\documentclass[fleqn,10pt]{wlscirep}
\usepackage[utf8]{inputenc}
\usepackage[T1]{fontenc}
\usepackage{amsmath,amssymb,amsfonts,amsthm,dsfont}
\usepackage[libertine,vvarbb]{newtxmath}
\usepackage{bm}
\usepackage{tabularx}
\usepackage[linesnumbered,ruled,vlined,boxed]{algorithm2e}
\usepackage{enumitem}
\usepackage{bibunits}
\usepackage{lscape}
\usepackage{longtable}
\usepackage{lineno}
\usepackage{multirow}
\usepackage{float}
\usepackage[most]{tcolorbox}

\title{SPIDER-WEB generates coding algorithms with superior error tolerance and real-time information retrieval capacity}

\author[1,2,3,5,$\dag$]{Haoling Zhang}
\author[4,$\dag$]{Zhaojun Lan}
\author[1,2]{Wenwei Zhang}
\author[1,2,5]{Xun Xu}
\author[1,2,3,5]{Zhi Ping}
\author[6,*]{Yiwei Zhang}
\author[1,2,3,5,*]{Yue Shen}
\affil[1]{BGI Research-Shenzhen, BGI, Shenzhen, 518083, China}
\affil[2]{Guangdong Provincial Key Laboratory of Genome Read and Write, BGI-Shenzhen, Shenzhen, 518120, China}
\affil[3]{George Church Institute of Regenesis, BGI-Shenzhen, Shenzhen, 518120, China}
\affil[4]{School of Mathematical Sciences, Capital Normal University, Beijing 100048, China}
\affil[5]{Shenzhen Institute of Synthetic Biology, Shenzhen Institutes of Advanced Technology, Chinese Academy of Sciences, Shenzhen 518055, China}
\affil[6]{School of Cyber Science and Technology, Shandong University, Qingdao, Shandong 266237, China}

\affil[*]{Corresponding authors: Yiwei Zhang (ywzhang@sdu.edu.cn) and Yue Shen (shenyue@genomics.cn)}
\affil[$\dag$]{These authors contributed equally to this work}

\begin{abstract}
DNA has been considered a promising medium for storing digital information. 
As an essential step in the DNA-based data storage workflow, coding algorithms are responsible to implement functions including bit-to-base transcoding, error correction, etc. 
In previous studies, these functions are normally realized by introducing multiple algorithms. 
Here, we report a graph-based architecture, named SPIDER-WEB, providing an all-in-one coding solution by generating customized algorithms automatically. 
SPIDERWEB is able to correct a maximum of 4\% edit errors in the DNA sequences including substitution and insertion/deletion (indel), with only 5.5\% redundant symbols.
Since no DNA sequence pretreatment is required for the correcting and decoding processes, SPIDER-WEB offers the function of real-time information retrieval, which is 305.08 times faster than the speed of single-molecule sequencing techniques. 
Our retrieval process can improve 2 orders of magnitude faster compared to the conventional one under megabyte-level data and can be scalable to fit exabyte-level data.
Therefore, SPIDER-WEB holds the potential to improve the practicability in large-scale data storage applications.
\end{abstract}

\begin{document}

\flushbottom
\maketitle
\thispagestyle{empty}

\rfoot {\thepage\ / \getpagerefnumber{sec:reference}}

\begin{bibunit}[naturemag]
\section*{Introduction}\label{sec:introduction}
The total amount of data increases exponentially as a result of the rapid development of human society. 
DNA molecule, as a candidate for storage medium with great potential~\cite{church2012next}, has drawn much attention due to its incredible storage capacity. 
The general workflow of DNA-based data storage includes: first, transforming the binary digital data into quaternary DNA sequences by specific coding algorithms~\cite{ping2019carbon}; second, synthesizing DNA molecules by chemical or enzymatic techniques to physically store the corresponding information; third, sequencing the DNA molecules and using the coding algorithm to retrieve the stored information. 
As the first and also the last step of the workflow, the coding algorithm plays an essential role in the effective transformation between binary messages and DNA sequences.

Early efforts of coding algorithm development aim to improve the biocompatibility of produced DNA sequences with DNA synthesis and sequencing techniques~\cite{kosuri2014large,treangen2012repetitive}. 
Specifically, most early established coding algorithms prohibited the generation of DNA sequences with long single-nucleotide repeats and extreme GC content~\cite{church2012next,goldman2013towards,grass2015robust,blawat2016forward,bornholt2017toward}. 
In a recent work, researchers further improved the biocompatibility by developing a coding algorithm based on the validity screening procedure to cope with more complicated constraints such as minimum free energy while reducing logical redundancy (i.e. the redundancy introduced at the coding-level for error tolerance)~\cite{ping2022towards}.
Considering the technological limitation during the DNA writing and reading process, errors typically occur in approximately 0.1-1.0\% of bases in the data coding DNA~\cite{press2020hedges}.
Such errors are the key confounding factors for lossless data retrieval. 
As a conventional solution, error correcting codes such as Hamming codes~\cite{hamming1950error}, Reed-Solomon codes~\cite{reed1960polynomial}, and Low-Density Parity-Check codes~\cite{gallager1962low} have been used together with the coding algorithms for the practice of DNA-based data storage~\cite{grass2015robust,chen2021artificial}. 
With the introduction of suitable logical redundancy, these error correcting codes could identify a certain amount of substitutions~\cite{ping2022towards}. 
However, errors such as insertion and deletion (indel) would not be able to be dealt with using conventional error correction codes. 
Although multiple sequence alignment can reduce the appearance of indel errors to some extent, it does not adapt to larger-scale applications beyond proof-of-principle experiments because of its high time cost~\cite{press2020hedges}. 
To address this issue, an algorithm named HEDGES is developed using a hash function and greedy exhaustive search to correct most indel and substitution errors. 
The probability depends on the introduction of logical redundancy. 
Specifically, to achieve a correcting performance at 2\% edit errors (including substitution, insertion and deletion), it requires at least 84.6\% logical redundancy~\cite{press2020hedges}. 
Nevertheless, such a search strategy makes the correcting process reach exponential computational complexity. 
Although the design of penalty coefficients can reduce such a computational complexity to a certain extent~\cite{press2020hedges,welzel2023dna}, it still introduces challenges to its practicability in large-scale applications.

Here we report a graph-based architecture ``SPIDER-WEB'', which is designed to automatically generate compatible coding algorithms for digital data to be stored with a built-in error correcting function. 
As a key and initial step of the architecture, SPIDER-WEB deposits required constraints into a customized directed graph and transforms this graph into a coding algorithm. 
For the generated coding algorithms, encoding and decoding processes are both executed by vertex transition in their corresponding graph. 
The restricted state transition path, i.e. generated DNA sequences, is also beneficial to error correction, which is achieved by local exhaustive reverse search and path-checking mechanism in our study. 
We have demonstrated that SPIDER-WEB can correct 4\% edit errors with the introduction of only 5.5\% logical redundancy.
In addition, compared with the greedy exhaustive search, our mechanism reduces the frequency of vertex access by thousands of times, leading to a significant decrease of computational complexity.
Therefore, SPIDER-WEB reaches the correcting speed of $137,288.36$ nucleotides per second, which is $305.09$ times faster than the maximum theoretical sequencing speed of single-molecule sequencing techniques~\cite{rang2018squiggle}. 
In addition, different from the conventional retrieval process (i.e. clustering $\rightarrow$ alignment $\rightarrow$ decoding $\rightarrow$ correcting), SPIDER-WEB provides a non-blocking process~\cite{marccais2011fast} based on the frequency statistics of corrected DNA sequences, which further accelerate data retrieval.
Without clustering and alignment pretreatments, such end-to-end retrieval can save operations to nearly one trillionth of that of the conventional retrieval process at the exabyte scale.
With its real-time information retrieval capacity, SPIDER-WEB offers the opportunity for large-scale data storage application scenarios in the near future.

\section*{Results}\label{sec:result}
\subsection*{Overview of SPIDER-WEB architecture}
\begin{figure}[H]
    \begin{center}
    \includegraphics[width=1\columnwidth]{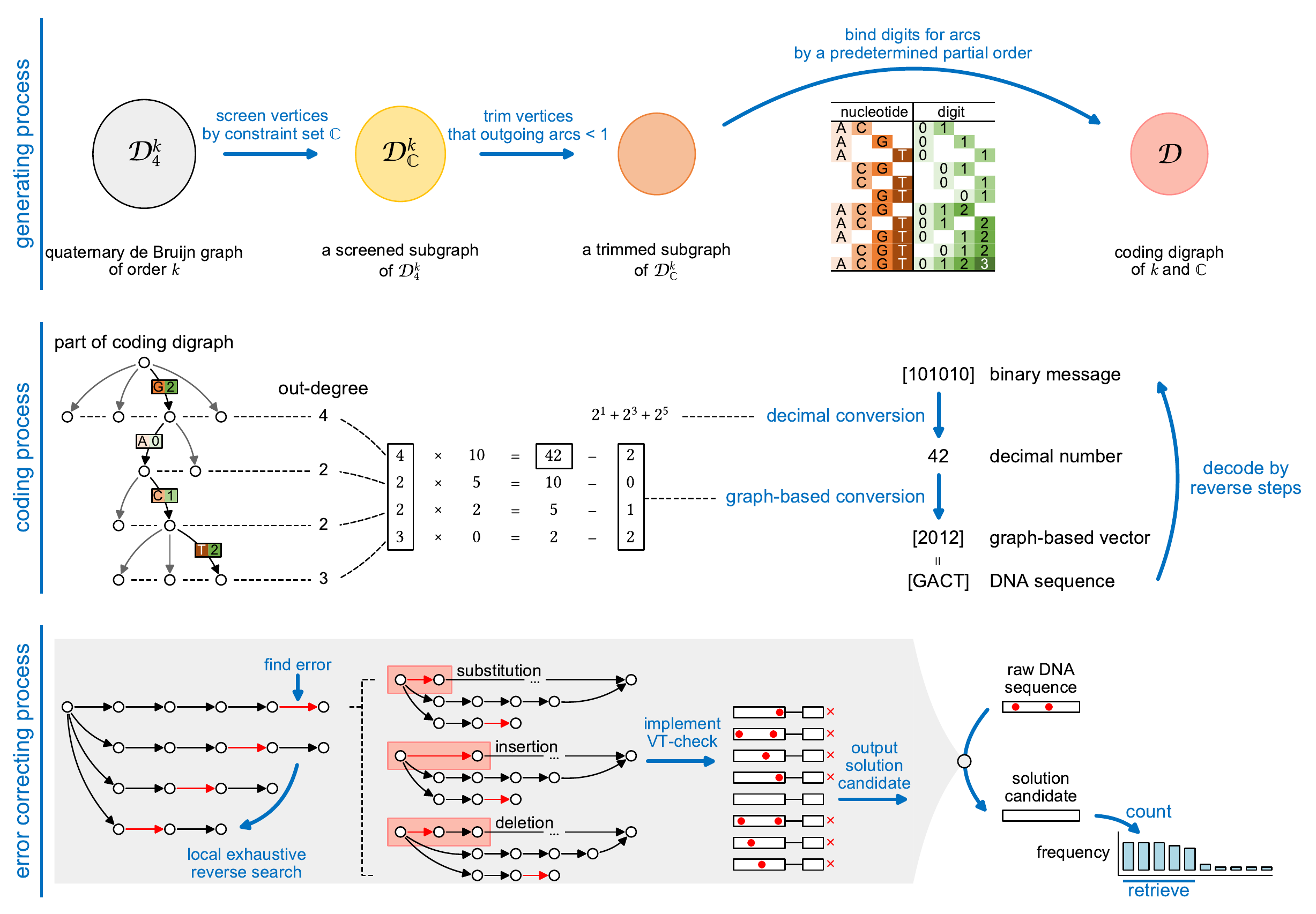}
    \caption{
    \textbf{Illustration of the SPIDER-WEB architecture.}
    This architecture can perform three processes: 
    (1) generate graph-based coding algorithms; 
    (2) encode binary messages as DNA sequences or decode DNA sequences as binary messages through the generated coding algorithm;
    (3) repair DNA sequences by path-based error correcting and further provide a real-time information retrieval based on frequency priority.}
    \label{fig:design}
\end{center}
\end{figure}

SPIDER-WEB architecture provides two basic functions including bit-to-nucleotide transcoding and error correcting, based on graph theory.
As Figure~\ref{fig:design}a shows, SPIDER-WEB generates the required bit-base transcoding algorithms by the following steps within 2 minutes (Appendix~\ref{sec:different_constraints}) before further applications: 
[1] select an observed length $k$ and generate all possible DNA $k$-mers, or vertices; 
[2] remove invalid DNA $k$-mers based on specified regional constraints; 
[3] form a directed de Bruijn graph using the rest $k$-mers;
[4] trim partial vertices to ensure that every vertex in the graph has a out-degree larger than $1$,
[5] bind edges to binary digits and convert this directed graph to a coding algorithm (hereinafter referred to as a coding digraph).
The information densities of the generated coding digraphs are calculated to be close to their theoretical upper bound (Appendix~\ref{sec:different_constraints}).

To encode binary message into DNA sequence through the coding digraph, a walking path is performed on the graph. 
As Figure~\ref{fig:design}b shows, in a standard coding digraph, traversing a directed edge from one vertex to the follow-up vertex yields the last nucleotide of the follow-up vertex and the bound digit of the edge (refer to Figure 2b). 
Therefore, assembling a DNA sequence that meets the specified constraints can be achieved by connecting the last nucleotides of the accessed vertices in sequence. 
Additionally, the bound digits acquired during this process can be utilized for a graph-based number-base conversion, resulting in the assembly of a binary message. 
By walking paths in the coding digraph, an effective mapping between binary messages and DNA sequences can be established.

Moreover, due to the constraints in selecting paths, SPIDER-WEB possesses an efficient and rapid correction method, namely path-based error correcting (Figure~\ref{fig:design}c and Figure~\ref{fig:repair}a). 
This approach adjusts paths that do not comply with the established constraints to their most similar restricted state transition paths, treating them as potential solutions.
The corresponding DNA sequence is then corrected probabilistically.
Additionally, path sieving is conducted using a modified version of the Varshamov-Tenengolts code, called VT-check~\cite{varvsamov1965code}. 
This process significantly improves the probability of obtaining a unique and correct solution candidate, thereby meeting the desired sequence correction requirement.
Furthermore, taking advantage of the ``multi-copy'' property of DNA molecules, a small number of incorrect solution candidates obtained through path-based error correction, or false positive sequences, can be screened based on their low frequency. 

\subsection*{Establishing a high-compatible coding digraph}
To be compatible with supporting techniques as much as possible, here we introduced a set of regionalized constraints (defined in Appendix~\ref{sec:local_formulation}) to establish the coding digraph. 
The commonly-used supporting techniques are DNA synthesis, polymerase chain reaction (PCR) amplification, DNA sequencing and preservation, which all require sequence constraints to some extent.
First of all, for DNA synthesis techniques, we set the homopolymer run-length as $2$ for further adapting enzymatic DNA synthesis~\cite{shafir2021sequence} beyond the previous constraint design~\cite{press2020hedges}.
And then, to improve the success rate of PCR amplification techniques~\cite{benita2003regionalized}, the regionalized GC content is set as exactly $50\%$.
Next, we forbade undesired motifs that may cause a high error rate in different sequencing techniques including Illumina sequencer (GGC)~\cite{press2020hedges} and ONT sequencer (AGA, GAG, CTC and TCT)~\cite{wick2019performance}.
Finally, we have reduced sequence features that are not conducive to preservation.
Basically, the preservation is divided into in \textit{vitro} and in \textit{vivo}, the compatibility requirements of the latter one are extremely complex.
For in \textit{vivo} preservation, these produced DNA molecules cannot affect the normal life cycle of the host, nor can they be broken by enzymes in the host.
Therefore, we treat start/stop codons (ATG, GTG, TTG, TAG, TAA and TGA)~\cite{dong2020dna} and restriction enzyme sites (GCT, GACGC, CAGCAG, GATATC, GGTACC, CTGCAG, GAGCTC,  GTCGAC, AGTACT, ACTAGT, GCATGC, AGGCCT and TCTAGA)~\cite{arber1969dna} as undesired motifs and prevent them from appearing in the produced DNA sequences.

The high-compatible coding digraph is generated according to the following steps. 
We initially had $4^{10}=1,048,576$ DNA fragments as vertices, based on the observed length of $10$ for better rounding. 
After being screened by the aforementioned constraints, the number of valid vertices decreased to $7,788$. 
To ensure reliable encoding, we applied a minimum out-degree threshold of $1$, which led to the removal of vertices without outgoing vertices. 
This trimming process reduced the number of vertices and obtained the resulting coding digraph contained $4,937$ vertices, with $1,741$ vertices having an out-degree of $2$ and the remaining vertices having an out-degree of $1$. 
The theoretical information density of the digraph reached $0.581$, as determined by the capacity approximator presented in Appendix~\ref{sec:math_proof}. 
The overall process of generating this coding digraph took $16.33$ seconds, including $15.03$ seconds for screening, $1.08$ seconds for trimming, and $0.22$ seconds for binding.

\subsection*{Correcting performance verification for a single DNA sequence}
\begin{figure}[htbp]
    \begin{center}
    \includegraphics[width=1\columnwidth]{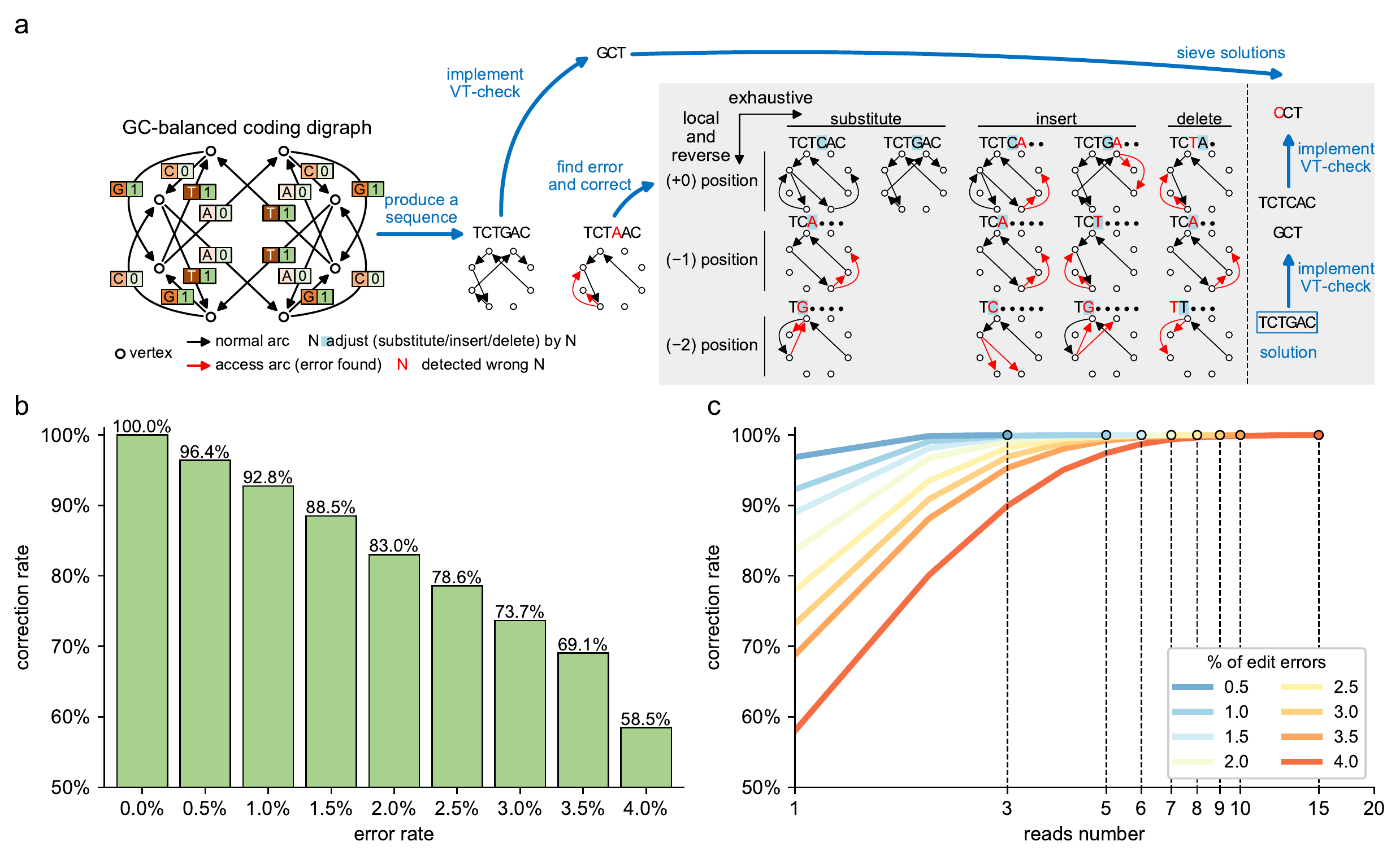}
    \caption{
    \textbf{Correcting performances of SPIDER-WEB for a single sequence.}
    (a) a simple case of path-based error correcting using a GC-balanced coding digraph. 
    (b) the correction rate under different error rates. 
    (c) the minimum reads number required to support full confidence that the sequence with the highest frequency is the correct sequence.}
    \label{fig:repair}
    \end{center}
\end{figure}

Using the aforementioned high-compatible coding digraph, we generate $10,000$ random samples (i.e. DNA sequence of length 200 nt) to evaluate the correcting performance of SPIDER-WEB.
For simplicity, we assume equal occurrence probability for substitutions, insertions and deletions with total error probability per nucleotide and errors occurred at random positions in the obtained DNA sequence~\cite{press2020hedges}. 
Based on these conditions, the rate of correcting all the errors under different error rates is shown in Figure~\ref{fig:repair}b.
When there is 0.5\% error (i.e. one edit error in a 200nt-long sequence), SPIDER-WEB can perform $96.4\%$ true positive correction, tying well with previous proofs of VT code on single error correction~\cite{varvsamov1965code}. 
When facing more edit errors, this correction rate decline with identical parameter settings as expected.
Even if the error rate of DNA sequences reaches $4\%$, which is the recent real raw error rate of single-molecule sequencing techniques~\cite{marx2023method}, the correction rate is still close to $60\%$.

To further investigate the impact of error rates on its detection/correction capability, we analyze other relative key factors during the correction process.
As a variable directly related to the correction rate, with the increase of error rates ($0.5\%$ to $4.0\%$), the rate of error detection is also decreasing (from $96.53\%$ to $76.67\%$; see Figure~\ref{fig:detection}).
It reveals that, after introducing more errors, a correct DNA sequence can be detected as another false-positive DNA sequence that satisfies the establish constraints and passes the verification of VT code.
Meanwhile, as shown in Figure~\ref{fig:repair_reduce}, these false-positive DNA sequences also affect the number of solution candidates.
It implies that the multiplicity of false positive sequences is one of the fundamental reasons for the decrease of correction rate.

Although the scoring mechanism can be used to further screen the solution candidate set, the ingenious penalty design for different error situations requires a large amount of prior knowledge.
Intriguingly, with the appropriate sequence copies, we can assume the correct sequence as the sequence with the highest frequency of occurrence in order to filter the false-positive ones.
Since the molecule product of each DNA sequence contains multiple copies in general, we can utilize this ``multi-copy'' feature to screen the false-positive sequences according to the low frequency.
To investigate the minimum sequence copies (as referred to reads number below) to support the above hypothesis, i.e. correct sequence with the highest frequency of occurrence, we performed $10,000$ random tests with above error rates.
As shown in Figure~\ref{fig:repair}b, to ensure the reliability of this hypothesis, the reads number of each sequence needs to be not smaller than $3$, $5$, $7$, $9$ and $15$ for $0.5\%$, $1.0\%$, $2.0\%$, $3.0\%$ and $4.0\%$ of edit errors respectively.
Therefore, after utilizing few sequence copies, the DNA sequence with highest frequency can be used as the target DNA sequence.

\subsection*{Demonstration of pretreatment-free retrieval mechanism}
Since the frequency statistics strategy for a single sequence is effective, we have the opportunity to retrieve information from the raw sequencing data without clustering and alignment pretreatments under the scale of the sequence population.
Based on the experience provided by Figure~\ref{fig:repair}b, it is natural to assume that the average frequency of all correct DNA sequences should be significantly greater than the average number of all the false positive DNA sequences with the proper reads number.
To apply to the pretreatment-free retrieval mechanism, the frequency statistics strategy needs a little modification.
After correcting all the DNA sequences from raw sequencing data, these corrected DNA sequences are graded (score $=$ frequency).
If $n$ DNA sequences are synthesized, only the top $n$ DNA sequences with highest score are retrieved for decoding.

\begin{figure}[htbp]
    \begin{center}
    \includegraphics[width=1\columnwidth]{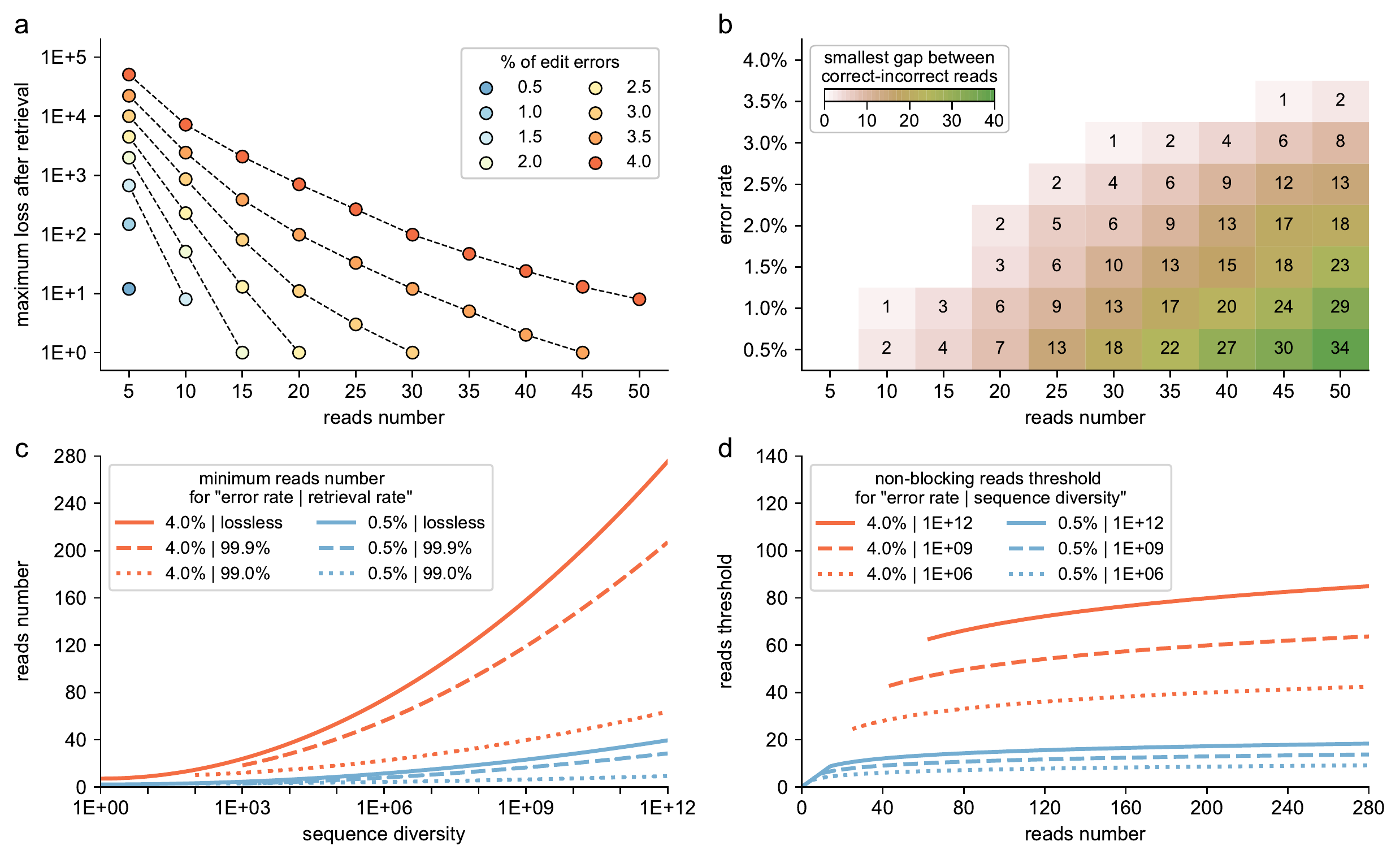}
    \caption{
    \textbf{Pretreatment-free retrieval mechanism performance of SPIDER-WEB.}
    Let the sequence diversity to be $10^6$ ($= 13.85$ megabyte data), (a) reports the maximum sequence loss after retrieval by the frequency statistics strategy under different error rates and different reads numbers; and (b) provides the smallest gap between correct and incorrect sequences.
    After intervening with the sequence diversity, (c) and (d) approximate the minimum reads number for lossless retrieval and non-blocking reads threshold for non-blocking retrieval respectively based on symbolic regression (Appendix~\ref{sec:symbolic_regression}).
    Raw data used to fit (c) and (d) are reported in Figure~\ref{fig:min_reads} and Figure~\ref{fig:min_threshold}.}
    \label{fig:pretreatment}
    \end{center}
\end{figure}

For the sequence population, different DNA sequences may obtain the same false-positive sequence after correcting, which may make the frequency of false-positive (or incorrect) sequences slightly increase.
Therefore, to find the limit of this expanded frequency statistics strategy, we create a large-scale experiment to detect the worst situation.
We first control the sequence diversity to explore the impact of different error rates and reads numbers on the retrieval rate.
The the sequence diversity is set as $10^6$ to approach the data size demonstrated by the previous wet experiments~\cite{erlich2017dna,press2020hedges,ping2022towards,song2022robust}.
Through $100$ random parallel experiments, Figure~\ref{fig:pretreatment}a reports the maximum loss after retrieval (i.e. number of incorrect retrieved sequences) by the frequency statistics strategy under different error rates and different reads numbers with the uniform readout distribution~\cite{rashtchian2017clustering}.
For error rates less than or equal to $1\%$, the retrieved $10^6$ DNA sequences still contain incorrect sequences when the reads number is $5$.
The maximum loss is $12$ for $0.5\%$ errors and $149$ for $1.0\%$ errors.
Consistent with our concern, with this reads number, the minimum frequency of the correct DNA sequence may be less than or equal to the maximum frequency of the incorrect sequence, thus partial correct sequences and incorrect sequences is indistinguishable.
Nevertheless, as shown in Figure~\ref{fig:pretreatment}b, the smallest gap between correct and incorrect DNA sequences can be widened when we increase the reads number, so that all retrieved DNA sequences are the correct sequences as expected.
Even in the case of high error rate, we can reach the expected retrieval rate by increasing the reads number.
Under $4.0\%$ error rate, the maximum loss are decreased from $50,747$ to $8$ as we increase the reads number from $5$ to $50$.
All in all, the frequency statistics strategy can work at the scale of sequence population, and we only need to adjust the reads number to achieve lossless retrieval.

To further investigate the influence of sequence diversity on our frequency statistics strategy, we intervened in the value of sequence diversity (i.e. $10^1$, $10^2$, $10^3$, $10^4$ and $10^5$).
We collect the maximum sequence loss for each sequence diversity scale with $100$ random parallel experiments and approximate minimum reads number required to reach the given retrieval rate under different sequence diversity via the symbolic regression method~\cite{keren2023computational}.
Figure~\ref{fig:pretreatment}c shows the fitting curve under different retrieval rates at the solvable bound of error rates.
For the lower bound of error rate (i.e. $0.5\%$), when the reads number is greater than or equal to $40$, terabyte-scale digital data can be retrieved losslessly.
Under $4.0\%$ errors, although the reads number needs to greater than or equal to $276$ for lossless retrieving terabyte-scale digital data, it can be reduced to $64$ for $99.0\%$ information retrieval.
Since the digital data that can reach the terabyte-scale is usually the context-correlation data such as videos, the incorrect part can be repaired through context correlation~\cite{tassano2020fastdvdnet} in the case of retrieving most of correct information.

Furthermore, our frequency statistics strategy can provide a non-blocking mechanism to cope with potential excessive memory consumption.
Following the results given by Figure~\ref{fig:pretreatment}b, for a given reads number, we can directly perform the decoding processing of a corrected DNA sequence without waiting for all corrected DNA sequences to be sorted when the frequency of this corrected DNA sequence exceeds the maximum frequency of incorrect DNA sequences.
Here we also complete the intervention experiment consistent with the above and observe the change of the maximum frequency of the incorrect sequence.
As shown in Figure~\ref{fig:pretreatment}d, the growth of the maximum frequency of the incorrect sequence (or non-blocking reads threshold) is far slower than that of the given reads number in the case of arbitrary error rate and sequence diversity.
For $0.5\%$ error rate, no matter how diverse the sequence in the population is, if the frequency of a corrected DNA sequence exceeds $20$, this DNA sequence can be directly decoded.
Although the error rate can affect the non-blocking reads threshold, in the case of high reads number (i.e. $\geq 152$), the non-blocking reads threshold is usually less than half of the given reads number.

\subsection*{Capacity for real-time information retrieval}
\begin{figure}[H]
    \begin{center}
    \includegraphics[width=1\columnwidth]{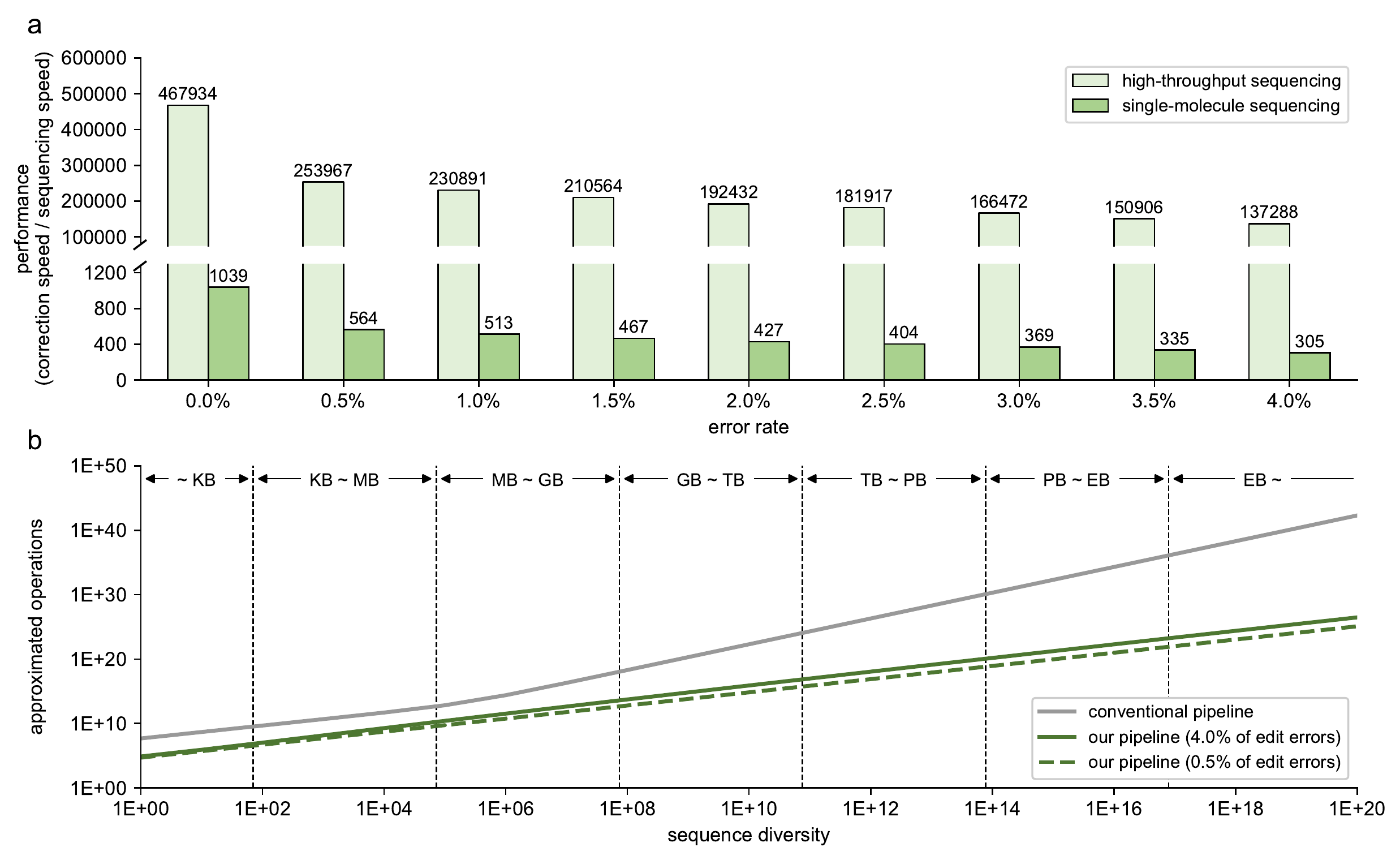}
    \caption{\textbf{Retrieval timeliness of SPIDER-WEB.}
    (a) describes the difference between average correction speeds under different error rates and the sequencing speeds of different sequencing platforms.
    Here, the theoretical maximum output of single-molecule sequencing techniques is set as $450$ nucleotides per second~\cite{rang2018squiggle} and that of high-throughput sequencing is set as $1$ nucleotide per second.
    (b) approximates the difference of operations between our process and the conventional retrieval process.
    The detailed operation approximation is reported in Appendix~\ref{sec:time_complexity}.}
    \label{fig:realtime}
    \end{center}
\end{figure}

Although DNA-based data storage is a storage technique for cold data, too long retrieval time from raw sequencing data to digital data may greatly hinder its commercial value.
Here, we use $10^5$ random samples to calculate the average correction speed per DNA sequence (length $= 200$ nt). 
Although the error rate can affect the correction time, our average correction time is one hundred-thousandth of high-throughput sequencing techniques and one percent of single-molecule sequencing techniques even at $4.0\%$ error rate (see Figure~\ref{fig:realtime}a).
This proves that the runtime of our corrector is negligible for any sequencing platform.

Since there is a significant difference between our process (correcting $\rightarrow$ sorting $\rightarrow$ decoding) and the conventional retrieval process (clustering $\rightarrow$ alignment $\rightarrow$ decoding $\rightarrow$ correcting), we further analyze the operation difference between two processes under the retrieval perspective. 
Theoretically, let $N$ be the sequence diversity, the computational complexity of our process is $O(N \times \log^2 N)$, which is far below $O(N^2)$ of the conventional retrieval process (see Appendix~\ref{sec:time_complexity}).
Based on the approximated average computational complexity, Figure~\ref{fig:realtime}b provides operations of two processes under different sequence diversities. 
It is assumed that the time required for operation in different steps of the process is consistent.
The runtime of our process takes hundredths of that of the conventional process at proof-of-concept data retrieval (megabyte-level).
When the retrieved data reaches exabyte-level, the number of our operations might be reduced by $12$ orders of magnitude.

\section*{Discussion}
SPIDER-WEB allows the coding algorithm to be produced under biochemical constraints without artificial design.
Since the generated hidden mapping connections do not need to retain mathematical elegance, it accepts more varied constraints while the information densities of its created coding algorithms are getting closer to their theoretical limits.
Meanwhile, unlike the random incorporation strategy~\cite{erlich2017dna,ping2022towards}, the specificity of input information can hardly affect the performance of generated coding algorithms, leading to a higher stability of the generated coding algorithms compared to the existing well-known algorithms. 
Hence, SPIDER-WEB can also bring a benchmark and beneficial supplement for follow-up artificial design of coding algorithms.
In a sense, when the information density of the designed algorithm is higher than that of the corresponding generated algorithm, such a design would be effective. 

In addition to the limitation of DNA synthesis cost, it is obvious that the commercialization of DNA-based data storage is also affected by the accuracy and runtime of information retrieval.
With the more digital data stored by DNA-based techniques, the higher the possibility of retrieving large-scale data at one time.
Unfortunately, the conventional retrieval process did not support information retrieval with volume exceeding gigabyte level data.
Under the condition of ensuring at most $4.0\%$ error tolerance, SPIDER-WEB can be ``painlessly'' embedded in various sequencing platforms, taking less than one percent of original runtime.
In addition, SPIDER-WEB provides an end-to-end retrieval mechanism, compressing the computational complexity of overall retrieval process from $O(N^2)$ to $O(N \times \log^2 N)$, where $N$ is the sequence diversity.
When retrieving exabyte-level information in the near future, the estimated runtime can be reduced by $12$ orders of magnitude compared with the conventional retrieval process, thus providing a novel vision to meet the demand of real-time correcting and paving the way for a commercial use of DNA-based data storage.

The aforementioned two aspects report the superiority of the SPIDER-WEB architecture. 
However, it is worth noting that there are also some issues to be further investigated.
Prospectively, various regional constraints could be expanded into various global constraints. 
By doing so, we might have the opportunity to include the toxicity, stability and expected structure of DNA sequences into consideration.
Besides, the generated coding algorithm with fixed-length could be further designed to accurately control the length of the produced DNA sequences.
Further, regarding one of the requirements of storage management, the risk of information eavesdropping and tampering should be avoided.
In addition to disturbing the modular operation in the encoding process to resist ciphertext-only attacks, for example, adjusting the predetermined partial order or trimming arcs that are finally retained, we could investigate whether there are potential methods to deal with various types of attack (Appendix~\ref{sec:protection}).

\section*{Methods}\label{sec:methods}
To explain the following processes, we first introduce some graph-theoretic terminologies. 
A digraph $\mathcal{D} = (\mathcal{V}, \mathcal{A})$ consists of a set of vertices $\mathcal{V}$ and a set of arcs (or directed edges) $\mathcal{A}$. 
Each arc from the vertex $\bm{u}$ to the vertex $\bm{v}$ is denoted as $(\bm{u},\bm{v})$, where $\bm{u}$ is called the initial vertex and $\bm{v}$ is called the terminal of the arc $(\bm{u},\bm{v})$. 
The number of arcs with $\bm{u}$ as initial vertex is called the out-degree of $\bm{u}$, denoted as $\text{deg}^{+}_\mathcal{D}(\bm{u})$.
In DNA-based data storage we always consider the quaternary alphabet $\{\text{A,C,G,T}\}$. 
Given positive integers $k$, the $4$-ary de Bruijn graph of order $k$, denoted as $\mathcal{D}^k_4$, is a digraph whose vertex set is $\mathcal{V}=\{\text{A,C,G,T}\}^k$. 
For any two vertices $\bm{u} = (\bm{u}[1], \bm{u}[2], \dots,\bm{u}[k])$ and $\bm{v} = (\bm{v}[1],\bm{v}[2],\dots,\bm{v}[k])$, there is an arc $(\bm{u},\bm{v})$ in $\mathcal{D}^k_4$ if and only if $\bm{u}[i+1]=\bm{v}[i]$ for $1\le i \le k-1$. 
It is routine to check that every vertex in $\mathcal{D}^k_4$ has out-degree $4$.

\subsection*{Algorithm generating}
The constraints considered in this paper include the homopolymer run-length constraints, the regionalized GC content constraints, and occasionally a third kind of constraint forbidding undesired motifs (mathematical formulations of these constraints are listed in Appendix~\ref{sec:local_formulation}). 
Following a standard approach in constrained coding theory, we consider a state-transition digraph $\mathcal{D}^k_{\mathbb{C}}$, which is a subgraph of the de Bruijn graph $\mathcal{D}^k_4$ induced by the vertices which satisfy the constraint set $\mathbb{C}$. 
The digraph $\mathcal{D}^k_{\mathbb{C}}$ characterizes the constraints in the sense that every quaternary codeword satisfying the constraints can be represented as a directed path in $\mathcal{D}^k_{\mathbb{C}}$, and vice versa.

In SPIDER-WEB, the initializing step is to build an underlying digraph $\mathcal{D}$ for further implementations. 
The coding algorithm starts with a screening process by deleting vertices that correspond to sub-sequences of length $k$ violating one or more constraints from $\mathbb{C}$.
After obtaining the digraph $\mathcal{D}^k_{\mathbb{C}}$ in step 1, it goes on to several recursive steps to further trim the vertex set, guaranteeing that the final output $\mathcal{D}$ has minimum out-degree of at least 1, to prevent that the encoding process stops halfway~\cite{press2020hedges}.

After constructing $\mathcal{D}$, we move onto a binding process, which binds each arc of $\mathcal{D}$
with a digit.
For each arc in $\mathcal{D}$ from $\bm{u}$ to $\bm{v}$, where $\bm{u}[1:k - 1] = \bm{v}[2:k]$, we call it a $v[k]$-arc, $\bm{v}[k] \in \{\text{A},\text{C},\text{G},\text{T}\}$
Each vertex has at most $\epsilon \leq 4$ outgoing arcs with distinct labels and we bind these arcs with digits $\{0,1,\dots,\epsilon-1\}$ according to a predetermined partial order $\text{A} < \text{C} < \text{G} < \text{T}$, as Figure~\ref{fig:design}. 
For example, if a vertex has three outgoing arcs with labels (like $\{\text{A},\text{C},\text{T}\}$), then (counting from zero) the $\text{A}$-arc is the $0$-th one, the $\text{C}$-arc is the $1$-st, and the $\text{T}$-arc is the $2$-nd.
These arcs are further labeled to as $\{\text{A}|0\}$-arc, $\{\text{C}|1\}$-arc, and $\{\text{T}|2\}$-arc, called label arc.
After obtaining label arcs, the coding digraph $\mathcal{D}$ is generated (detailed in Algorithm~\ref{al:generating}).

\subsection*{Graph-based encoding}
Here, $\mathcal{D}$ will lead to a coding algorithm $\mathcal{E}$, which encodes $\bm{x}$, a binary message of length $k$, into a DNA sequence $\bm{y} = \mathcal{E}(\bm{x})$, and guarantees that $\bm{y}$ satisfies the constraints $\mathbb{C}$ (as Algorithm~\ref{al:encoding}).

An example of graph-based encoding is shown in Figure~\ref{fig:design}.
Based on the state transition, a piece of information has undergone $3$ transformations from a binary message to a DNA sequence.
Initially, the binary message $[101010]$ is converted into the decimal number $42 = 2^1 + 2^3 + 2^5$.
Such a decimal number is further disassembled into a graph-based vector through $\mathcal{D}$.
The value in each position of this vector depends on the out-degree of the vertex currently being accessed.
In this example, the out-degree of the virtual vertex is 4, the graph-based vector can obtain 2 in the first cycle because $42 = 4 \times 10 + 2$, and the decimal number becomes $10$.
The next cycle could be $10 = 2 \times 5 + 0$.
The cycle ends until the decimal number becomes 0 and we obtain a graph-based vector $[2012]$.
Reversely, based on the fixed mapping in each arc, $[2012]$ can be decoded as $[\text{GACT}]$.

\subsection*{Path-based error correcting}
Probabilistically, DNA sequences obtained from the sequencing process may contain one or more errors, where the error type could be insertions, deletions, or substitutions. 
Many works on error correcting codes against one type of error or a combination of two types of error have been done by coding theorists.
However, if we consider the combination of all three types (which is referred to as an edit error), up till now there are only codes against one edit error. 
Constructing codes against multiple edit errors is still a challenging task in coding theory.

At first sight, it is tempting to directly bring some known codes against one edit error into our SPIDER-WEB framework for error correction. 
Nevertheless, there are two disadvantages. 
Firstly, a proportion of the codewords may not satisfy the established constraint set and thus cannot be used. 
Secondly, when multiple errors occur (which is quite common), the decoding process for codes against only one edit error will no longer help.

Due to the structure of $\mathcal{D}$, our encoding scheme naturally brings some robustness against errors in the following sense.
If there is no error then the output of our coding algorithm must be a directed path on $\mathcal{D}$, which is shown in Figure~\ref{fig:repair}a. 
Otherwise, when errors occur then with high probability the output erroneous sequence will no longer be a valid directed path on $\mathcal{D}$.
Therefore, we may check the validity of an output sequence by tracing the directed path on $\mathcal{D}$. 
Whenever we are at some vertex in $\mathcal{D}$ but fail to find the suitable label arc corresponding to the next nucleotide of the sequence, then we know that the nucleotide of current and/or previous positions is wrong.
Thus, such errors can be detected timely.

Once an error is spotted at the $j$-th position, we apply a local exhaustive reverse search method for error correction (Figure~\ref{fig:repair}a).
Here, ``local'' and ``reverse'' mean that we check backward from the $j$-th to the $(j-k)$-th position (in some sense according to a decreasing order of the error probability).
When checking a particular position $i$, $j-k \leq i \leq j$, we guess the type of error and try these 3 adjustment types: (1) substitute the nucleotide in its current position with another nucleotide; (2) insert a nucleotide between the $(i-1)$-th position and the $i$-th position; and (3) delete the nucleotide in the $i$-th position.
Conservatively, all adjustment types need to be tested in each position of the local range (which accounts for ``exhaustive''). 
The number of adjustments includes at most 3 substitutions, 4 insertions and 1 deletion. 
The precise number depends on the outgoing arcs of the current working vertex in $\mathcal{D}$. 
If an adjustment results in a sequence that (partly) corresponds to a valid directed path on $\mathcal{D}$, we consider it as a possible correct candidate. 
This process works if there is only one edit error. 
In case there are multiple edit errors, as long as the errors are separated enough, we may repair the errors sequentially.

It might happen that the local exhaustive reverse search provides massive candidates, especially when the constraints are not too strict. 
We further apply a sieving method by tools from coding theory, in order to significantly reduce the number of candidates or even find a unique solution (Figure~\ref{fig:repair_reduce}). 
The trick is to apply an idea similar to the Varshamov-Tenengolts code and its variations, which play an important role in the current study of error correcting codes against deletions or insertions. 
What we do is to provide a ``salt-protected'' DNA sequence of length $(k + 1)$ as a suffix for each codeword. 
By salt-protection~\cite{blawat2016forward,press2020hedges}, the suffix is guaranteed to be correct. 
It stores some information known as the check value $\bm{y}_\text{check}$
for a codeword $\bm{y}$ (Appendix~\ref{sec:modular_operation}).

The process of our correction strategy (Algorithm~\ref{al:repairing}) is defined as $\bm{Y}_\mathrm{repair} = \mathcal{C}(\bm{y}_\mathrm{wrong}, \bm{y}_\mathrm{check}, \mathcal{D})$.
First, we use the local exhaustive reverse search to find a list of candidates. 
Then for each candidate we compute its check values and see if it matches the correct values stored in the salt-protected suffix. 
While it is still possible that the algorithms may fail when facing multiple and dense errors, it behaves well in our simulation (see Figure~\ref{fig:repair}b).

\subsection*{Pretreatment-free retrieval mechanism}
Since the molecule product of each DNA sequence contains multiple copies in general, it is a common process to complete clustering and alignment from sequencing data successively before obtaining DNA sequence for decoding.
With the further development of DNA storage, the data capacity to be stored will increase significantly (like terabyte-level data instead of megabyte-level data for proof-of-concept).
However, the existing clustering methods or alignment methods have high computational complexity, and thus cannot be compatible with exabyte-level of data~\cite{press2020hedges,qu2022clover}.
As an end-to-end solution, we apply the path-based error correcting to all the raw sequencing sequences without the above pretreatments. 
Each raw sequence leads to a candidate set and the correct original information sequence should belong to most of the candidate sets. 
Thus we can order all the candidate sequences according to their frequency and pick the most frequent one as our decoding output.

\section*{Code Availability}
Kernel codes of SPIDER-WEB are exhibited in the ``dsw'' folder of the GitHub repository (\url{https://github.com/HaolingZHANG/DNASpiderWeb}).
The usage examples and customized suggestions of each class or interface are described on the ReadtheDocs website (\url{https://dnaspiderweb.readthedocs.io/en/latest/}).

Further, this repository also includes the process codes of all simulation experiments in its ``experiments'' folder.
All experiments were performed with a random seed of $2021$ in Windows 10 environment and Python 3.7.3, including an Intel$^\circledR$ Core$^\text{TM}$ i7-4710MQ CPU and $16$GB DDR3 RAM.

\section*{Data Availability}
The supplementary data underlying this article is shared on the above-mentioned GitHub repository.

\section*{Acknowledgments}
This work was supported by the National Key Research and Development Program of China (No. 2020YFA0712100), the National Natural Science Foundation of China (Nos. 12001323, 32101182, and 12231014), the Shandong Provincial Natural Science Foundation (No. ZR2021YQ46), the Guangdong Provincial Key Laboratory of Genome Read and Write (No. 2017B030301011). 
This work was also supported by China National GeneBank and George Church Institute of Regenesis, BGI-Shenzhen, China.

We thank Prof. George Church from Harvard University for constructive discussions on experiment design; 
Dr. Ryan Wick from Monash University for valuable discussions on Oxford Nanopore sequencing; 
Prof. Qingshan Jiang from Chinese Academy of Sciences, Dr. Guangyu Zhou from Harvard University, Prof. Eitan Yaakobi from Technion, Dr. Hector Zenil from University of Oxford, and Dr. Chao-Han Huck Yang from Georgia Tech for useful comments on the manuscript.

\section*{Author contributions}
H.Z designed the graph-based encoding;
H.Z and Z.L designed the local exhaustive reverse search and Varshamov-Tenengolts path check;
H.Z designed the pretreatment-free retrieval mechanism;
Z.L and Y.Z constructed the capacity approximator and completed the mathematical proofs;
H.Z implemented Python codes with optimized data structure and algorithms;
H.Z, Z.P and Y.S prepared the figures, tables and data;
H.Z and Z.L mainly drafted the manuscript;
Y.Z, Z.P, Y.S, W.Z, and X.X revised the manuscript;
Y.Z and Y.S supervised the study jointly.
All authors read and approved the final manuscript.

\putbib[reference]\label{sec:reference}
\end{bibunit}

\newpage
\appendix

\begin{bibunit}[naturemag]
\renewcommand\thesection{\Alph{section}}
\renewcommand\thefigure{S\arabic{figure}}   
\renewcommand\thetable{S\arabic{table}}  
\setcounter{figure}{0}
\setcounter{table}{0}
\thispagestyle{empty}
\rfoot{\thepage\ / \getpagerefnumber{sec:addition_reference}}

\noindent \textbf{\Huge Supplementary file of}

\vspace{0.5cm} 

\noindent{\LARGE SPIDER-WEB generates coding algorithms with superior error tolerance and real-time information retrieval capacity}
\vspace{0.5cm}

\noindent{Haoling Zhang$^\dag$, Zhaojun Lan$^\dag$, Wenwei Zhang, Xun Xu, Zhi Ping, Yiwei Zhang$^*$, Yue Shen$^*$}
\vspace{0.5cm}

\noindent{Correspondence to: ywzhang@sdu.edu.cn and shenyue@genomics.cn}
\vspace{1cm}

\noindent {\Large \textbf{Contents}}
\begin{table}[h]
\footnotesize
\resizebox{\linewidth}{!}{
\begin{tabular}{clcr}
\ref{sec:local_formulation} &  Mathematical formulation of the regional biochemical constraints & $\cdots$ & \getpagerefnumber{sec:local_formulation} / \getpagerefnumber{sec:addition_reference} \\
\ref{sec:math_proof} & Mathematical proof of capacity approximation under specific biochemical constraints & $\cdots$& \getpagerefnumber{sec:math_proof} / \getpagerefnumber{sec:addition_reference}\\
\ref{sec:modular_operation} & Modular operation setting of the salt-protected suffix storing check values & $\cdots$& \getpagerefnumber{sec:modular_operation} / \getpagerefnumber{sec:addition_reference} \\
\ref{sec:different_constraints} & Performance evaluation under different constraints & $\cdots$& \getpagerefnumber{sec:different_constraints} / \getpagerefnumber{sec:addition_reference} \\
 & 1 \quad Generating performance  &  \\
 & 2 \quad Coding performance & \\
\ref{sec:time_complexity} & Computational complexity deduction of end-to-end retrieval process & $\cdots$& \getpagerefnumber{sec:time_complexity} / \getpagerefnumber{sec:addition_reference} \\
 & 1 \quad Pretreatment-free retrieval process & \\
 & 2 \quad Conventional retrieval process & \\
\ref{sec:symbolic_regression} & Equation fitting based on the symbolic regression & $\cdots$ &  \getpagerefnumber{sec:symbolic_regression} / \getpagerefnumber{sec:addition_reference} \\
\ref{sec:protection} & Variations and capabilities for privacy protection & $\cdots$ &   \getpagerefnumber{sec:protection} / \getpagerefnumber{sec:addition_reference}  \\
\ref{sec:optimization} & Software optimization and reliability analysis & $\cdots$ &   \getpagerefnumber{sec:optimization} / \getpagerefnumber{sec:addition_reference}  \\
 & 1 \quad Index definition in the programming \\
 & 2 \quad Representation of digraph & \\
 & 3 \quad Use of rapid search in digraph & \\
 & 4 \quad Approximation of largest eigenvalue & \\
 & 5 \quad Reliability analysis of capacity approximation & \\
\ref{sec:pseudo_code} & Detailed pseudo code for proposed algorithms in this work & $\cdots$ &   \getpagerefnumber{sec:pseudo_code} / \getpagerefnumber{sec:addition_reference} \\
 & 1 \quad Initializing step to build the digraph of the coding algorithm &  \\
 & 2 \quad Encoding binary message through graph-based coding algorithm &  \\
 & 3 \quad Repair DNA sequence through path-based error correcting &  \\
 & 4 \quad Approximate the capacity &  \\
\ref{sec:addition_information} & Supporting figures and tables & $\cdots$ &  \getpagerefnumber{sec:addition_information} / \getpagerefnumber{sec:addition_reference} \\
 & 1 \quad Representative regional biochemical constraint sets (Table~\ref{tab:constraint_sets}) &  \\
 & 2 \quad Generation runtime of SPIDER-WEB under different biochemical constraint sets (Figure~\ref{fig:generation_runtime}) &  \\
 & 3 \quad Remaining vertex number after screening process of SPIDER-WEB (Figure~\ref{fig:screening_vertex}) &  \\
 & 4 \quad Remaining vertex number during trimming process of SPIDER-WEB (Table~\ref{tab:trimming_vertex}) &  \\
 & 5 \quad Adjustable parameter list of well-established coding algorithms in the coding task (Table~\ref{tab:parameter_1}) &  \\
 & 6 \quad Adjustable parameter list of SPIDER-WEB in the coding task (Table~\ref{tab:parameter_2}) &  \\
 & 7 \quad Coding performance of different coding algorithms (Table~\ref{tab:coding_performance}) &  \\
 & 8 \quad Standard deviation of different coding algorithm performances (Table~\ref{tab:stability}) &  \\
 & 9 \quad Decoding success probability of repetitive patterns under the error-free retrieval (Table~\ref{tab:hedges_rate}) &  \\
 & 10 \; Information density of graph-based coding algorithms versus their corresponding approximated capacity (Figure~\ref{fig:code_rate}) &  \\
 & 11 \; Detection rate and correction rate under different error rates (Figure~\ref{fig:detection}) &  \\
 & 12 \; Effect of Varshamov-Tenengolts path check (Figure~\ref{fig:repair_reduce}) &  \\
 & 13 \; Raw experiment data of minimum reads number under different error rates and retrieval rates (Figure~\ref{fig:min_reads}) &  \\
 & 14 \; Raw experiment data of maximum frequency of incorrect reads under different error rates and sequence diversities (Figure~\ref{fig:min_threshold}) &  \\
 & 15 \; Average vertex access frequency of local and global search under different error rates (Table~\ref{tab:runtime}) &  \\
 & 16 \; Combination size under different constraints based on DNA sequence lengths (Figure~\ref{fig:protection}) &  \\
 & 17 \; Relative error statistics of the capacity approximation using random digraphs (Figure~\ref{fig:capacity_evaluation}) &  \\
\hyperlink{page.32}{K} & References & $\cdots$ &  35 / \getpagerefnumber{sec:addition_reference} \\
\end{tabular}}
\end{table}

\clearpage

\setcounter{page}{1}

\section{Mathematical formulation of the regional biochemical constraints}
\label{sec:local_formulation}
According to the previous works~\cite{church2012next,goldman2013towards,grass2015robust,blawat2016forward,erlich2017dna,gabrys2020locally}, two kinds of biochemical constraints are widely investigated: the homopolymer run-length constraint and the regionalized GC content constraint.

In a DNA sequence, a homopolymer run refers to a maximal consecutive sub-sequence of the same symbol and the number of nucleotides in each run is called its run-length. 
For example, $v=\mathrm{AAAATTCGG}$ contains four runs with run-lengths as 4,2,1,2 accordingly.
Typically in DNA-based data storage long runs should be avoided. 
A homopolymer run-length constraint is of the form ``the maximal run-length is at most $h$".

Another constraint, known as the regionalized GC content, requires that the GC-ratio in any consecutive sub-sequence of length $k$ is bounded within an interval. 
The parameter $k$ is called the observed length and usually $k\ge h$. Given $0 \leq \epsilon_1 \leq \epsilon_2 \leq 1$ and the observed length $k$, a regionalized GC content constraint is of the form ``in any consecutive sub-sequence of length $k$, the sum of the numbers of G and C is within the interval $[\epsilon_1 \times k,\epsilon_2 \times k]$.''

Additionally, we sometimes come across a third kind of constraint regarding undesired motifs. 
Such motifs usually have a serious impact on specific biochemical operations or storage environments. Let $\Theta$ be a set of undesired motifs. 
A DNA sequence $v$ is $\Theta$-free if each motif in $\Theta$ does not appear in $v$ as a consecutive sub-sequence. 
Usually we only consider undesired motifs with length less than or equal to the observed length $k$. 

\clearpage

\section{Mathematical proof of capacity approximation under specific biochemical constraints}
\label{sec:math_proof}
Given a set of constraints $\mathbb{C}$, we want to encode binary messages into quaternary DNA sequences satisfying the constraints. 
In classical coding theory, we usually want the encoded codewords to be of a fixed length. 
That is, we want to encode binary messages from $\{0,1\}^m$ into quaternary DNA sequences in $\{\text{A,C,G,T}\}^n$. 
The efficiency of the code is characterized by the information density $r$, defined as $r=m\ /\ n$. 
The maximal information density is called the capacity of such codes. 
For example, if there are no constraints, then by a trivial mapping from $\{00,01,10,11\}$ to $\{\text{A,C,G,T}\}$, we have a code of rate $2$. 

However, due to the constraints $\mathbb{C}$ we must have a sacrifice on the information density. 
Let ${_4}\Sigma^n_{\mathbb{C}}$ be the set of DNA sequences in $\{\text{A,C,G,T}\}^n$ which satisfy the constraints $\mathbb{C}$. 
For any $m$, the maximal information density will be $m\ /\ n^\prime$ where $n^\prime$ is the least integer such that $|{_4}\Sigma^{n^\prime}_{\mathbb{C}}|\ge 2^m$. 
The precise computation of $|{_4}\Sigma^{n^\prime}_{\mathbb{C}}|$ is a difficult problem. 
In constrained coding theory, a standard way to asymptotically compute $|{_4}\Sigma^n_{\mathbb{C}}|$ relies on the celebrated Perron-Frobenius Theorem~\cite{perron1907theorie,frobenius1912matrizen} and the procedure is as follows. 
Given the constraints $\mathbb{C}$, consider its state-transition digraph $\mathcal{D}^k_{\mathbb{C}}$, where $k$ is chosen as the observed length in a regionalized GC content constraint from $\mathbb{C}$. 
In fact, such a graph is exactly the graph we mentioned in Algorithm~\ref{al:generating} after the screening process. 
As long as $\mathcal{D}^k_{\mathbb{C}}$ is strongly-connected, the spectral radius $\rho$ of the adjacency matrix of $\mathcal{D}^k_{\mathbb{C}}$ will provide the estimation
\begin{equation}
\label{eq:capacity}
\lim_{n \to \infty} |{_4}\Sigma^n_{\mathbb{C}}|\approx\rho^n,
\end{equation}
and thus the capacity is upper bounded by $\log_2 \rho$.

While coding theorists might focus on fixed-length coding algorithms with rate approaching $\log_2 \rho$, in this paper we consider variable-length codes. 
That is, our SPIDER-WEB coding algorithm might encode binary messages from $\{0,1\}^m$ into DNA sequences of variable lengths. 
For a comparison with the fixed-length model, for any variable-length coding algorithm $\mathcal{E}$, define $\overline{n}=\frac{1}{2^m} \sum_{x\in\{0,1\}^m} |\mathcal{E}(x)|$ to be the average length of the encoded codewords and the rate of $\mathcal{E}$ is defined as $r(\mathcal{E})=m\ /\ \overline{n}$.

Since fixed-length codes are special cases of variable-length codes, at first sight we might expect $r(\mathcal{E})$ to be larger than the capacity of fixed-length codes. 
However, a key observation is that $\log_2 \rho$ is still an upper bound of $r(\mathcal{E})$. 
Following this observation, we may compare the information density of SPIDER-WEB and the corresponding capacity result and it gives supportive evidences that our SPIDER-WEB coding algorithms indeed have good information densities. 
The rest of this subsection is devoted to the proof of the key observation: 

\begin{equation} 
\label{eq:upper_bound}
\lim_{m\rightarrow +\infty} r(\mathcal{E}) \leq \log_2{\rho},
\end{equation}

\begin{proof}[Proof of equation~\ref{eq:upper_bound}]

Let the set of binary messages be $\{0,1\}^m$ where $m$ can be arbitrarily large. Let $n^\prime$ be the smallest integer such that $|{_4}\Sigma^{n^\prime}_{\mathbb{C}}|\ge 2^m$. 
A fixed-length code will have rate $m\ /\ n^\prime$. 
A variable length code, however, could use codewords with smaller length. 
Without loss of generality, we assume in a greedy way that this variable length code uses codewords with length as small as possible. 
Divide all the $2^m$ codewords into sets $\mathcal{E}_1,\mathcal{E}_2,\dots,\mathcal{E}_{n^\prime}$ where $\mathcal{E}_i$ represents the set of codewords of length $i$. 
Then by definition $\overline{n}=\frac{1}{2^m}\sum_{i=1}^{n^\prime}i|\mathcal{E}_i|$.

Pick an auxiliary variable $0<\epsilon<1$, divide $\sum_{i=1}^{n^\prime}i|\mathcal{E}_i|$ into two parts and the computation proceeds as follows.
\begin{equation*}
\frac{1}{r(\mathcal{E})} = \frac{1}{m \times 2^m} \times \sum_{i=1}^{n^\prime} i|\mathcal{E}_i| = \frac{1}{m \times 2^m}  \times  \Big(\sum_{i=1}^{\frac{\epsilon \times m}{\log_2 \rho}} i|\mathcal{E}_i| + \sum_{i=\frac{\epsilon \times m}{\log_2 \rho} + 1}^{n^\prime} i|\mathcal{E}_i| \Big) \geq \frac{1}{m \times 2^m}  \times \Big(\sum_{i=1}^{\frac{\epsilon \times m}{\log_2 \rho}} i|\mathcal{E}_i| + \frac{\epsilon \times m}{\log_2 \rho}  \times \sum_{i=\frac{\epsilon \times m}{\log_2 \rho}+1}^{n^\prime} |\mathcal{E}_i|\Big).
\end{equation*}

By substituting $\sum_{i=\frac{\epsilon \times m}{\log_2 \rho}+1}^{n^\prime} |\mathcal{E}_i|=2^m-\sum_{i=1}^{\frac{\epsilon \times m}{\log_2 \rho}} |\mathcal{E}_i|$, we arrive at 
\begin{equation*}
\frac{1}{r(\mathcal{E})}\geq \frac{1}{m \times 2^m} \times  \sum_{i=1}^{\frac{\epsilon \times m}{\log_2 \rho}} i|\mathcal{E}_i| + \frac{\epsilon}{2^m  \times \log_2 \rho} \times  \Big(2^m-\sum_{i=1}^{\frac{\epsilon \times m}{\log_2 \rho}} |\mathcal{E}_i|\Big) 
= \frac{\epsilon}{\log_2 \rho} + \frac{1}{2^m} \times \sum_{i=1}^{\frac{\epsilon \times m}{\log_2 \rho}} (\frac{i}{m}-\frac{\epsilon}{\log_2 \rho})|\mathcal{E}_i|.
\end{equation*}

Note that for $1\leq i\leq \frac{\epsilon \times m}{\log_2 {\rho}}$ we have $\frac{i}{m}-\frac{\epsilon}{\log_2 \rho}\leq 0$.
According to Equation~(\ref{eq:capacity}), $|\mathcal{E}_i|\le |{_4}\Sigma^i_{\mathbb{C}}| \approx \rho^i$. 
Then we proceed as follows.
\begin{equation*}
\frac{1}{r(\mathcal{E})}\geq \frac{\epsilon}{\log_2 \rho} + \frac{1}{2^m} \times \sum_{i=1}^{\frac{\epsilon  \times m}{\log_2 \rho}} (\frac{i}{m}-\frac{\epsilon}{\log_2 \rho})|\mathcal{E}_i| \geq \frac{\epsilon}{\log_2 \rho} + \frac{1}{2^m} \times \sum_{i=1}^{\frac{\epsilon \times m}{\log_2 \rho}} (\frac{i}{m}-\frac{\epsilon}{\log_2 \rho})\rho^i.
\end{equation*}

Denote $\Delta=\sum_{i=1}^{\frac{\epsilon \times m}{\log_2 \rho}} (\frac{i}{m}-\frac{\epsilon}{\log_2 \rho})\rho^i$. 
By calculating the difference between $\rho \times \Delta$ and $\Delta$ we have
\begin{equation*}
(\rho-1) \times \Delta 
= -\frac{1}{m} \times \sum_{i=1}^{\frac{\epsilon \times m}{\log_2 \rho}} \rho^i + \frac{\epsilon \times \rho}{\log_2 \rho} \geq - \frac{1}{m}  \times \frac{\rho \times (\rho^{\frac{\epsilon \times m}{\log_2 \rho}}-1)}{\rho-1}
= -\frac{\rho \times 2^{\epsilon \times m}}{m \times (\rho-1)}+\frac{\rho}{m \times (\rho-1)}.
\end{equation*}

Then finally we have 
\begin{equation*}
\frac{1}{r(\mathcal{E})}\geq \frac{\epsilon}{\log_2 \rho} - \frac{\rho \times 2^{\epsilon \times m}}{2^m  \times m \times (\rho-1)^2} + \frac{\rho}{2^m  \times m \times (\rho-1)^2 }.
\end{equation*}

When taking $m\rightarrow \infty$, the second and the third term on the right hand side approach 0 and thus we have proven 
\begin{equation*}
\lim_{m\rightarrow \infty} \frac{1}{r(\mathcal{E})} \geq  \frac{\epsilon}{\log_2 \rho}.
\end{equation*}
Since the parameter $\epsilon$ can be chosen arbitrarily close to $1$, we have finally proven 
\begin{equation*}
\lim_{m\rightarrow \infty} r(\mathcal{E}) \leq \log_2\rho.
\end{equation*}
\end{proof}

\clearpage

\section{Modular operation setting of the salt-protected suffix storing check values}
\label{sec:modular_operation}
Consider an arbitrary map between $\{\text{A,C,G,T}\}$ and $\{0,1,2,3\}$, say 
\begin{equation*}
   \text{A} \leftrightarrow 0, \text{C} \leftrightarrow 1, \text{G} \leftrightarrow 2, \text{T} \leftrightarrow 3.
\end{equation*}

For a quaternary DNA sequence $\bm{y}=(\bm{y}[1], \bm{y}[2], \cdots, \bm{y}[n])$, define its signature as 
\begin{equation*}
   \text{sig}(\bm{y})=(\bm{x}[1], \bm{x}[2], \cdots, \bm{x}[n-1]), 
\end{equation*}
where
\begin{equation*}
\bm{x}[i] = \left\{
\begin{array}{cc}
1 & \bm{y}[i+1]\geq \bm{y}[i] \\
0 & \bm{y}[i+1]<\bm{y}[i]
\end{array}
\right.
\end{equation*}

The check value $\bm{y}_\mathrm{check}$ for a codeword $\bm{y}$ is of length $k+1$ and consists of two parts. The first entry of $\bm{y}_\mathrm{check}$ is the value
\begin{equation*}
\sum_{i=1}^n \bm{y}[i] \pmod{4}.
\end{equation*}
The last $k$ entries is the quaternary expression of the number
\begin{equation*}
\sum_{i=1}^{n-1} i \times \bm{x}[i] \pmod{4^k}.
\end{equation*}

The idea behind these check values is the celebrated Varshamov-Tenengolts code~\cite{varvsamov1965code,tenengolts1984nonbinary}. 
If there is only one deletion or one insertion, then from the check values we may precisely correct the error. In this paper we do not directly apply the decoding algorithms of VT codes or their variations, since they are vulnerable against multiple errors. 
Instead, we only use these check values as a sieving method, to further check the correctness for each candidate sequence arisen from our search method.

\clearpage

\section{Performance evaluation under different constraints}
\label{sec:different_constraints}
\subsection{Generation performance}
Here, we introduce $12$ representative regional constraint sets (Table~\ref{tab:constraint_sets}), to analyze the influence of biochemical constraints on Generating performance of SPIDER-WEB. 

\begin{figure}[ht]
    \begin{center}
    \includegraphics[width=1\columnwidth]{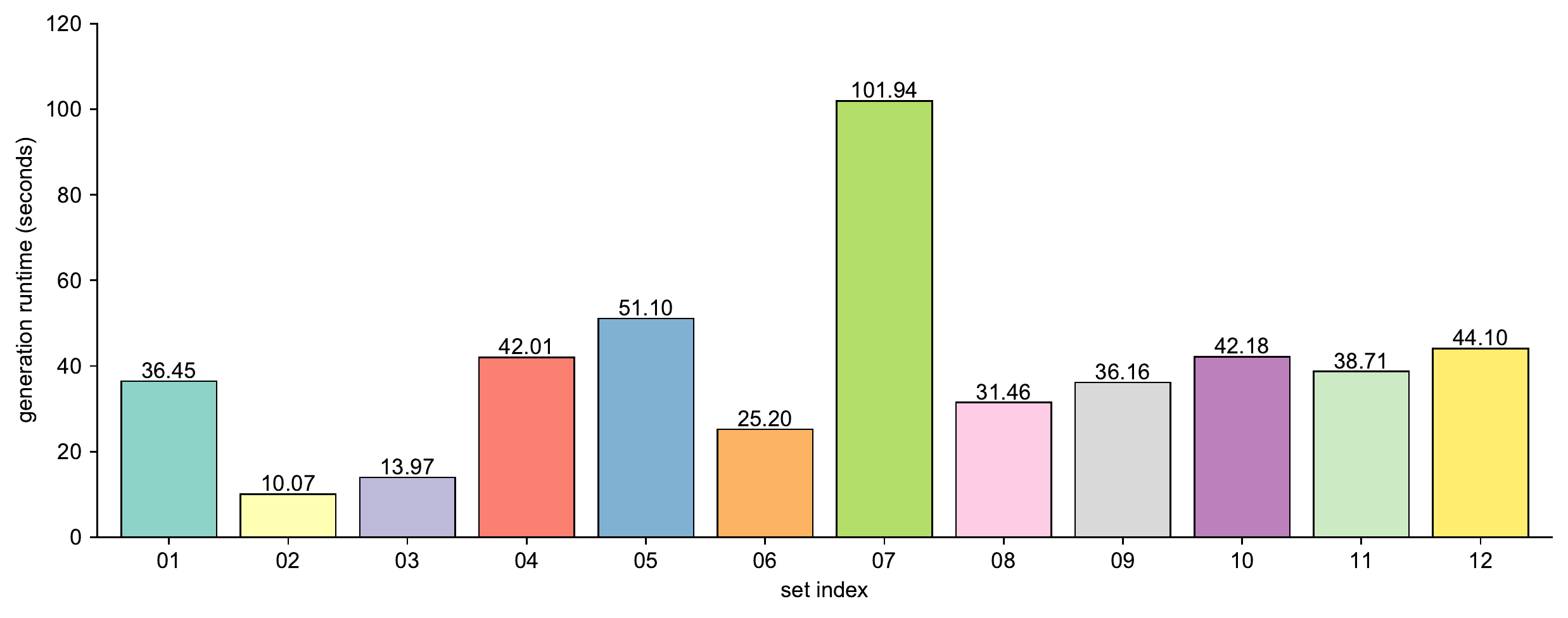}
    \caption{\textbf{Generation runtime of SPIDER-WEB under different biochemical constraint sets in Table~\ref{tab:constraint_sets}.}
    To reasonably validate the runtime, all statements related to Monitor should be removed in the scripts.}
    \label{fig:generation_runtime}
    \end{center}
\end{figure}

To generate algorithms under these constraints, SPIDER-WEB took an average of $39.45$ seconds ($101.94$ seconds at most; see Figure~\ref{fig:generation_runtime}).
Considering that these constraint combinations cover different tendencies, we can boldly speculate that the time of algorithm generation is minute level.
Potentially, it can effectively replace artificial algorithm design without introducing longer or more complex constraints (e.g. minimum free energy~\cite{ping2022towards}, hazardous genes~\cite{puzis2020increased}, distances for hybridization~\cite{bee2021molecular,tomek2021promiscuous}).

The tendentiousness of constraints, i.e. undesired motifs lead to asymmetry of valid DNA $k$-mers, has an impact on the generation process even the coding process (Figure~\ref{fig:code_rate}). 
For the trimming process, it can lead to more rounds of arc trimming (Table~\ref{tab:trimming_vertex}).
Worse, it may cause the coding digraph to fail to generate.

\subsection{Coding performance}
Three early-established coding algorithms might possess the ability to deal with various regional biochemical constraints recently: DNA Fountain~\cite{erlich2017dna}, Yin-Yang Code~\cite{ping2022towards}, and HEDGES~\cite{press2020hedges}. 
To investigate the practical information density (see below) for aforementioned coding algorithms, $100$ groups of above-mentioned algorithm parameter (Table~\ref{tab:parameter_1} and \ref{tab:parameter_2}) and $100$ groups of bit matrix are randomly introduced. 

In order to be close to the actual situation, each randomized bit matrix contains $72$ kilobytes (include $8$ kilobytes index range and $64$ kilobytes payload range) for the simulated information density experiment.
Based on the principle of fair comparison, the bit matrix has different dimensions for different algorithms, making the length of DNA sequences obtained by different algorithms as consistent as possible. 
The practical information density can be easily calculated, that is,
\begin{equation*}
\frac{\text{nucleotide number of the matrix}}{\text{payload range in the matrix}}.
\end{equation*}

Table~\ref{tab:coding_performance} demonstrates that coding algorithm produced by SPIDER-WEB performs better in encoding tasks than the other investigated algorithms under different constraints in Table~\ref{tab:constraint_sets}. 
For constraint set 02 -- 12, the lower bound of information density of our proposed algorithm is larger than the upper bound of DNA Fountain, Yin-Yang Code, and HEDGES. 
It can be considered the optimal selection to against arbitrary local biochemical constraints. 
Although HEDGES can provide higher coding performance in the partial constraint sets through customized patterns, the related settings may lead to decoding failure because of \textit{de facto} erasures (see Table~\ref{tab:hedges_rate}). 
The highest coding performance with a certain success decoding rate ($\geq$ 10\%) for reported patterns of HEDGES is reported in Table~\ref{tab:coding_performance}.

Besides, algorithms proposed in this work provide more stable coding performance. 
Under the above 12 constraint sets, the standard deviation of generated algorithms is not exceeding $0.011$. 
On the contrary, as shown in Table~\ref{tab:stability}, screening-based methods are more susceptible to different constraints, parameter setups, and/or digital information patterns (either the information density is low or the standard deviation of information density is greater than $0.011$). 

\clearpage

\section{Computational complexity deduction of end-to-end retrieval process}
\label{sec:time_complexity}
\subsection{Pretreatment-free retrieval process}
The retrieval process of SPIDER-WEB includes three parts: correcting, sorting and decoding. 
Therefore, its computational complexity is the sum of the computational complexity of three parts.
For a retrieval task, the parameters of calculate computational complexity are as follow:
\begin{itemize}
\item $N$: sequences diversity;
\item $R$: reads number;
\item $L$: sequence length (set as $200$ in this work);
\item $E$: number of errors in sequence ($1$--$8$ for $0.5\%$--$4.0\%$ of edit errors);
\item $F$: vertex access frequency per sequence during the correcting process ($227.45$ for $0.5\%$ of edit errors and $405.28$ for $4.0\%$ of edit errors; detailed in Table~\ref{tab:runtime});
\item $C_e$: number of solution candidates per sequence after the local exhaustive reverse search ($1.3936$ for $0.5\%$ of edit errors and $16.2931$ for $4.0\%$ of edit errors; detailed in Table~\ref{fig:repair_reduce}).
\item $C_p$: number of solution candidates per sequence after the VT-check path sieving ($1.0004$ for $0.5\%$ of edit errors and $1.1908$ for $4.0\%$ of edit errors; detailed in Table~\ref{fig:repair_reduce}).
\item $O_{\mathcal{D}\rightarrow10}$: average operation numbers from a graph-based vector to a decimal value ($24275.83$ when $L$ equals $200$).
\item $O_{10\rightarrow2}$: average operation numbers from a decimal value to a binary message ($24712.86$ when $L$ equals $200$).
\end{itemize}
Considering the arbitrariness of DNA sequence length, the value range of float-point number cannot may not represent, we complete the number-base conversions through the addition, subtraction, multiplication and division of strings.
It ensures the security of decimal conversion, but also adds additional calculation steps.
When the length of DNA sequence is $200$, we have completed $100$ random experiments for each starting vertex in the coding digraph.

During the correcting part, SPIDER-WEB detects one error and then corrects one error (Figure~\ref{fig:repair}a).
Since there may be false-positive solution candidates in the correction results, 
these candidates are screened by the VT-check path sieving, the calculation times of which is $L$.
Therefore, the computational complexity of this part could be
\begin{equation*}
N \times R \times (F + C_e \times L).
\end{equation*}

Affected by false-positive solution candidates, the total number of sequences obtained is change from $N \times R$ increased to $N \times R \times C_p$.
During the sorting part, all obtained sequences should be counted in advance.
With the classical collection method in Python (i.e. ``collections.Counter''), the computational complexity of counting process is equals to that of hash search.
Since any distance factor~\cite{rashtchian2017clustering} is not considered in the counting operation, the operations for calculating a hash value of a sequence is $L$.
For the key-value dictionary, the number of counting operations have
\begin{equation*}
(N \times R \times C_p) \times L.
\end{equation*}
The implementation of listing the $N$ most common sequences in the above mentioned Python interface is Quick Sort algorithm.
Practically, we do not pay attention to the sequence of sequences at the same frequency.
Hence, the sorting part can be further optimized.
If we create an inverse mapping of the above dictionary, key is count and value is the list of sequences, the computational complexity of sorting can be reduced from at most $(N \times R \times C_p) \times \log_2(N \times R \times C_p)$ to at most $(N \times R \times C_p) + (N + R)$. 
Therefore, the computational complexity of this part could be
\begin{equation*}
(N \times R \times C_p) \times (L + 1) + (N + R)
\end{equation*}

According to the extraction rules after sorting, $N$ DNA sequences with the highest frequency are decoded. 
For each DNA sequence, its decoding process contains two number-base conversions: DNA sequence is converted into decimal value and then a binary message is allocated from the aforementioned decimal value.
To simplify, the computational complexity of this part is
\begin{equation*}
N \times (O_{\mathcal{D} \rightarrow 10} + O_{10 \rightarrow 2}).
\end{equation*}

To sum up, the overall computational complexity of our end-to-end retrieval could be
\begin{equation*}
\overbrace{N \times R \times (F + C_e \times L)}^{\text{correct}}+ \overbrace{(N \times R \times C_p) \times (L + 1) + (N + R)}^{\text{sort}} + \overbrace{N \times (O_{\mathcal{D}\rightarrow10} + O_{10\rightarrow2})}^{\text{decode}}.
\end{equation*}
Using the minimum reads number (obtained by Appendix~\ref{sec:symbolic_regression}), $R$ can be equivalent to $O(\log^2 N)$.
When the error rate $e$ and sequence length $L$ are regarded as constants, the overall computational complexity of this lossless end-to-end retrieval under sequence diversity $N$ approximates
\begin{equation*}
    O(N \log^2 N).
\end{equation*}
Introducing known statistical values into this computational complexity to approximate the average number of operations, we need $183.18 \times N \times \log_{10}^2 N + 50,508.16 \times N + 0.26 \times \log_{10}^2(N) + 2.15$ operations for $0.5\%$ of edit errors and $7,275.66 \times N \times \log_{10}^2 N + 76,702.78 \times N + 1.86 \times \log_{10}^2(N) + 7.10$ operations for $4.0\%$ of edit errors.

\subsection{Conventional retrieval process}
In the conventional retrieval process, the data retrieval process is composed of clustering, multiple sequence alignment, decoding and error correcting. 
Therefore, its computational complexity is the sum of the computational complexity of the four parts.
To better compare the two processes, we continue to use the above variables, those are
\begin{itemize}
\item $N$: sequences diversity;
\item $R$: reads number (set as $5$ in Erlich et al. work~\cite{erlich2017dna});
\item $L$: sequence length (set as $200$ nucleotides in this work).
\end{itemize}

For the clustering methods, Rashtchian et al. outperforms any methods that require ``cluster center number $\times$ total reads number'' time regime~\cite{rashtchian2017clustering}.
Here, ``cluster center number'' is $N$ and ``total reads number'' is $N \times R$.
When $R$ can ignore the timing of $N$ (as discussed in this work), the computational complexity of this part is
\begin{equation*}
N \times N \times L.
\end{equation*}

On the basis of clustering, $N \times R$ sequences will be divided into $N$ clusters of size $R$ ideally. 
After that, each high-confidence DNA sequence can be obtained by aligning all DNA sequences in a cluster.
In a cluster, the alignment method uses $R \times L^3$ operations to deal with $R$ sequences of length $L$~\cite{maiolo2018progressive}.
Therefore, the computational complexity of this part is
\begin{equation*}
N \times R \times L^3.
\end{equation*}

For any decoding strategy, its minimum computational complexity per DNA sequence is also $L$.
Therefore, the computational complexity of this part is
\begin{equation*}
N \times L.
\end{equation*}

Finally, assuming that there is no insertion and deletion error in this case, lossless retrieval can be corrected by using Reed-Solomon Code~\cite{reed1960polynomial,welcherror}, the computational complexity of this part is
\begin{equation*}
N \times L^3.
\end{equation*}

All in all, the overall computational complexity of conventional retrieval could be
\begin{equation*}
O(N^2)
\end{equation*}
With the setting values above, the operations can be approximated as $200 \times N^2 + 48,000,200 \times N$.

\clearpage

\section{Equation fitting based on the symbolic regression}
\label{sec:symbolic_regression}
To find the appropriate fitting curves, based on the test points, we use the well-established symbolic regression toolbox ``pySR'' to complete the equation fitting~\cite{cranmer2020discovering}.
The selected operators are: ``div'', ``mult'', ``plus'', ``sub'', ``neg'', ``square'', ``cube'', ``pow'', ``exp'' , ``log2'' and ``log10'' for finding a complex function rather than simple polynomial equations.
We train the above regression model based on the gradient descent with the $L_1$ loss function (achieving lower loss values than the mean squared error loss function in the regression optimization task~\cite{qi2020mean}).

The following is our final equations.
With the raw data in Figure~\ref{fig:min_reads}, the minimum reads number $R_m$ for different retrieval rates under different errors and different sequence diversities $N$ can be generated as 
\begin{equation*}
R_m \approx \left\{
\begin{array}{clc}
1.864 \times \log_{10}^2(N) + 7.100 & \text{retrieval rate} = \text{lossless} & \text{errors} = 8 \\
1.399 \times \log_{10}^2(N) + 5.684 & \text{retrieval rate} = 99.9\% & \text{errors} = 8 \\
0.384 \times \log_{10}^2(N) + 8.572 & \text{retrieval rate} = 99.0\% & \text{errors} = 8 \\
0.259 \times \log_{10}^2(N) + 2.147 & \text{retrieval rate} = \text{lossless} & \text{errors} = 1 \\
0.189 \times \log_{10}^2(N) + 1.181 & \text{retrieval rate} = 99.9\% & \text{errors} = 1 \\
0.046 \times \log_{10}^2(N) + 2.627 & \text{retrieval rate} = 99.0\% & \text{errors} = 1 \\
\end{array}
\right..
\end{equation*}
With the raw data in Figure~\ref{fig:min_threshold}, under different errors, the non-blocking threshold $\tau$ (or maximum frequency of incorrect reads) for different reads number ($R$) and different sequence diversities ($N$) can be generated as 
\begin{equation*}
\tau \approx \left\{
\begin{array}{cc}
\log_{10}(R + 1) \times \log_{10}(N)\ /\ 0.346 & \text{errors} = 8 \\
\log_{10}(R + 1) \times \log_{10}(N)\ /\ 1.601 & \text{errors} = 1 \\
\end{array}
\right..
\end{equation*}
Under $4.0\%$ of edit errors, When $R$ is relative low, i.e. less than $24$ for $10^6$ sequence diversity, $42$ for $10^9$ sequence diversity and $62$ for $10^{12}$ sequence diversity, $\tau$ is greater than $R$.
Since lossless retrieval cannot be achieved under above reads numbers, we do not draw this part of curve in Figure~\ref{fig:pretreatment}d.

\clearpage
\section{Variations and capabilities for privacy protection}
\label{sec:protection}
The coding digraph produced by the established constraints serves as the essential foundation and upholds the effectiveness in handling errors.
Considering the random errors introduced by supporting techniques, the coding digraph is considered to be directly negotiated by information owners rather than stored and transmitted through DNA molecules together with the encoded digital information.
In a sense, it can be regarded as the encryption key to information transmission.
However, eavesdroppers may crack the information contained in the DNA molecules after reconstructing the coding digraph from the obtained DNA sequences~\cite{compeau2011apply}. 
After evaluating the correctness of the reconstructed coding digraph, the information can be easily obtained through the normal correcting and decoding processes.

To address this defect, as the variation of graph-based encoding, the bit-to-base mapping of partial arcs in $\mathcal{D}$ can be shuffled.
For a common DNA sequence, two binary messages decoded from the default setting and shuffled setting contain the avalanche-level difference because of the extensive impact of modular operations.
Through this confounder, the combination size under different constraint sets is reported in Figure~\ref{fig:protection}.
Users can choose one of the combinations to save as a key.
This alleviates the cracking danger in information transmission of DNA-based data storage under the ciphertext-only attack.

\clearpage

\section{Software optimization and reliability analysis}
\label{sec:optimization}
For graph-related calculations, the adjacency matrix is conventionally used to represent a digraph.
For the observed length $k$ mentioned above, the shape of an adjacency matrix is $(4^k, 4^k)$.
When $k$ reaches $8$, the file size will achieve $16.0$ gigabytes with the integer format~\cite{zuras2008ieee}, which is unable to be allocated (both MATLAB and Python platforms).
Considering that such adjacency matrix is an extremely sparse matrix, only $4$ positions in each row ($4^k$) can be actually used.
Using the compressed matrix such as compressed sparse row can save memory space exponentially, but it brings trouble to massive matrix operation (e.g. singular value decomposition).
It implies that the computing time of capacity approximation, coding algorithm generation, and graph-based search may be intolerable.
Thus, in this work, we have completed some design and optimization at the software level to make the creation and calculation of huge digraphs possible.

\subsection{Index definition in the programming}
Here we declare that the index in this section is defined as the zero-based numbering~\cite{seed2012introduction}.
Therefore, the initial element of a vector ($\gamma$) is assigned the index $0$, denote as $\gamma[0]$.
Besides, Numpy package~\cite{van2011numpy} accepts negative indices for indexing from the end of the vector.
For example, we have a vector $\gamma= (1, 2, 3,\cdots, 10)_{1 \times 10}$, $\gamma[0] = 1$ and $\gamma[-1] = \gamma[9] = 10$.

\subsection{Representation of digraph}
Let $\text{A} = 0$, $\text{C} = 1$, $\text{G} = 2$, and $\text{T} = 3$, the index of a vertex in $\mathcal{V}^k_4$ can be defined as a decimal number:
\begin{equation*}
\text{index}(\bm{u}) = \sum_{p=1}^k \bm{u}[p] \times 4^{k - p}.
\end{equation*}
Set $i = \text{index}(\bm{u})$, then $\bm{u}$ is $i$-th vertex of $\mathcal{V}^k_4$, also denote as $\mathcal{V}^k_4[i]$.
Hence, based on the de Bruijn graph of order $k$, for any arc $(\mathcal{V}^k_4[i], \mathcal{V}^k_4[j])$, $i$ and $j$ must satisfy $i \bmod 4^{k - 1} = \lfloor j / 4 \rfloor$.

Practically, some arcs are trimmed during special processes (like Algorithm~\ref{al:generating}).
As the simplest design, a special compressed matrix with $4^k$ rows (for vertex indices) and $4$ columns (for outgoing arcs) can be constructed.
For this matrix, if the element of $x$-th row and $y$-th column is $1$, $\mathcal{D}$ represented by this matrix contains an arc from $x$-th vertex to $(x \times 4 + y) \bmod 4^k$ vertex; otherwise, $\mathcal{D}$ lacks this arc.  
However, this work involves massive matrix, graph, and tree operations.
The above matrix will perform modular operation every time the vertices are accessed in turn.
Hence, we further construct a special compressed matrix, named accessor ($\alpha$), containing $4^k$ rows and $4$ columns.
If $(\mathcal{V}^k_4[i], \mathcal{V}^k_4[j])$ is an arc of $\mathcal{A}$, $\alpha[i, j \bmod 4] = j$.
When accessing a location (or column) in the row, it is directly to obtain the follow-up row corresponding to current row and column.
Besides, the value of the remaining elements of $\alpha$ is set as $-1$.
For any related operation, it will be ignored.
In practice, we only need to allocate a vector of size $4^k + 1$, so that the operation of these irrelevant values takes place in the last position without affecting the calculation itself.
After calculations, by removing the last position, the expected vector will be obtained.

To illustrate the difference between adjacency matrix and accessor, we represent the digraph in Figure~\ref{fig:repair}a by the following two kinds of matrices:

\begin{equation*}
\left(
\begin{array}{cccccccccccccccc}
0 & 0 & 0 & 0 & 0 & 0 & 0 & 0 & 0 & 0 & 0 & 0 & 0 & 0 & 0 & 0 \\
0 & 0 & 0 & 0 & 1 & 0 & 0 & 1 & 0 & 0 & 0 & 0 & 0 & 0 & 0 & 0 \\
0 & 0 & 0 & 0 & 0 & 0 & 0 & 0 & 1 & 0 & 0 & 1 & 0 & 0 & 0 & 0 \\
0 & 0 & 0 & 0 & 0 & 0 & 0 & 0 & 0 & 0 & 0 & 0 & 0 & 0 & 0 & 0 \\
0 & 1 & 1 & 0 & 0 & 0 & 0 & 0 & 0 & 0 & 0 & 0 & 0 & 0 & 0 & 0 \\
0 & 0 & 0 & 0 & 0 & 0 & 0 & 0 & 0 & 0 & 0 & 0 & 0 & 0 & 0 & 0 \\
0 & 0 & 0 & 0 & 0 & 0 & 0 & 0 & 0 & 0 & 0 & 0 & 0 & 0 & 0 & 0 \\
0 & 0 & 0 & 0 & 0 & 0 & 0 & 0 & 0 & 0 & 0 & 0 & 0 & 1 & 1 & 0 \\
0 & 1 & 1 & 0 & 0 & 0 & 0 & 0 & 0 & 0 & 0 & 0 & 0 & 0 & 0 & 0 \\
0 & 0 & 0 & 0 & 0 & 0 & 0 & 0 & 0 & 0 & 0 & 0 & 0 & 0 & 0 & 0 \\
0 & 0 & 0 & 0 & 0 & 0 & 0 & 0 & 0 & 0 & 0 & 0 & 0 & 0 & 0 & 0 \\
0 & 0 & 0 & 0 & 0 & 0 & 0 & 0 & 0 & 0 & 0 & 0 & 0 & 1 & 1 & 0 \\
0 & 0 & 0 & 0 & 0 & 0 & 0 & 0 & 0 & 0 & 0 & 0 & 0 & 0 & 0 & 0 \\
0 & 0 & 0 & 0 & 1 & 0 & 0 & 1 & 0 & 0 & 0 & 0 & 0 & 0 & 0 & 0 \\
0 & 0 & 0 & 0 & 0 & 0 & 0 & 0 & 1 & 0 & 0 & 1 & 0 & 0 & 0 & 0 \\
0 & 0 & 0 & 0 & 0 & 0 & 0 & 0 & 0 & 0 & 0 & 0 & 0 & 0 & 0 & 0
\end{array}
\right)_{16 \times 16}
\iff
\left(
\begin{array}{rrrr}
-1 & -1 & -1 & -1  \\
4 & -1 & -1 & 7    \\
8 & -1 & -1 & 11   \\
-1 & -1 & -1 & -1  \\
-1 & 1 & 2 & -1    \\
-1 & -1 & -1 & -1  \\
-1 & -1 & -1 & -1  \\
-1 & 13 & 14 & -1  \\
-1 & 1 & 2 & -1    \\
-1 & -1 & -1 & -1  \\
-1 & -1 & -1 & -1  \\
-1 & 13 & 14 & -1  \\
-1 & -1 & -1 & -1  \\
4 & -1 & -1 & 7    \\
8 & -1 & -1 & 11   \\
-1 & -1 & -1 & -1
\end{array}
\right)_{16 \times 4}.
\end{equation*}
By doing so, the memory usage of digraph in this work can be reduced from $4^{k + k}$ to $4^{k + 1}$.
In our simulation experiments ($k = 10$), the allocated memory is decreased from $4.0$ terabytes to $16.0$ megabytes (262,144 times).

\subsection{Use of rapid search in digraph}
In the generation task of SPIDER-WEB, breadth first search~\cite{moore1959shortest} is used widely times (line 2 -- 5 of Algorithm~\ref{al:generating}). 
Meanwhile, in the repair task of SPIDER-WEB, a major repetitive operation (line 5 -- 8 of Algorithm~\ref{al:repairing})  is to detect whether the repaired DNA sequence is on the path that satisfies established biochemical constraints.
Thus, the optimization of search operation in digraph can significantly reduce the time required for not only graph generating but also path-based error correcting.

Through accessor, the search operation can be converted to matrix operation, we therefore construct a rapid search for layers and paths.
Here, we introduce stride $s$ to represent the stride from the root vertex to the leaf vertex, then provide below customized algorithms.
Normally, in the breadth-first search~\cite{cormen2009introduction}, the vertex acquisition of the next layer depends on the next vertex of each vertex in the current layer. 
Accessor considers parallelize the search of each layer, so as to speed up the search.

Taking $i$-th vertex ($\mathcal{V}^k_4[i]$) as the root vertex in the digraph $\mathcal{D}$, the initial vector can be $\gamma_0 = (0, 0, \cdots, 1, \cdots, 0, 0)_{1 \times 4^k + 1}$.
In this vector, only $i$-th element is $1$.
For the next layer (level $1$), we initialize $\gamma_1$ as an all-zero vector and fill the locations (represented by the value in $\alpha[i]$) of $\alpha[i]$ in $1$.
That is, 
\begin{equation*}
\gamma_1[\alpha[i,j]] = 1
\end{equation*}
where $j \in \{0, 1, 2, 3\}$.
Expansively, the $s$-layer vector can be defined through $s$ iterations. 
In each iteration, all the elements in the vector is initialized as $0$.
From $s$-th iteration to $(s+1)$-th iteration,
\begin{equation*}
\gamma_{s+1}[\alpha[i]] = \sum_{f=0}^3 \gamma_s[f \times 4^{k - 1} + \lfloor \frac{i}{4} \rfloor],
\end{equation*}
where $j$ is each location with the value of $1$ in $\gamma_s$.
Using the ``where'' function in Numpy package, the search process can be executed in parallel (from $4^k$ iterations to $4$ iterations).

\subsection{Approximation of largest eigenvalue}
As mentioned in Appendix~\ref{sec:math_proof}, the capacity under the considered biochemical constraints can be approximated based on the largest eigenvalue of its corresponding generated digraph.
Theoretically, the largest eigenvalue ($\rho$) can be approximated directly through the QR transformation~\cite{francis1961qr} (Q is an orthogonal matrix and R is a right triangular matrix). 
Considering that accessor is not suitable for matrix decomposition operations, such approximation method should be replaced by the power iteration~\cite{mises1929praktische}.
Hence, the largest eigenvalue (Algorithm~\ref{al:approximation}) can be approximated as:
\begin{equation*}
\rho = \lim_{t \to \infty}\max(\gamma_t),
\end{equation*}
where $\gamma_t$ is the eigenvector obtained from $t$-th iteration and $t$ approaches infinity.

As an initialized eigenvector, $\gamma_0$ can be an all-one vector or a random vector~\cite{kuczynski1992estimating}.
The size of eigenvectors (and normalized temporary vector, see below) is $4^k + 1$ and the last element is reset to $0$.
$\gamma_{i + 1}$ is calculated by $\gamma_i$ and $\alpha$ with the follow-up two steps.
The first step is to calculate a normalized temporary vector from $\gamma_i$, as
\begin{equation*}
\tau[\epsilon] = \frac{\gamma_i[\epsilon]}{\max(\gamma_i)},
\end{equation*}
where $\epsilon \in [0, 4^k]$.
And in the second step, for any index $\epsilon_1$ in $\gamma_{i+1}$,
\begin{equation*}
\gamma_{i+1}[\epsilon_1] =
\sum_{f=0}^3 \tau[\epsilon_2]
\end{equation*}
where
\begin{equation*}
\epsilon_2 = \alpha[f \times 4^{k - 1} + \lfloor \frac{\epsilon_1}{4} \rfloor, \epsilon_1 \bmod 4].    
\end{equation*}
Here, for $\epsilon_1$-th vertex, $f \times 4^{k - 1} + \lfloor \frac{\epsilon_1}{4} \rfloor$ refers to its former vertex.
And the decimal number of first nucleotide in this former vertex is $k$.
In practice, we cannot obtain the result when the iteration approaches infinity.
When an appropriate tolerance
\begin{equation*}
\frac{|\max(\gamma_{i}) - \max(\gamma_{i - 1})|}{\max(\gamma_{i - 1})} \leq 10^{-10}
\end{equation*}
is met~\cite{ford2014numerical}, we believe that there is no difference between the $i$-th largest eigenvalue and $\infty$-th largest eigenvalue.

\subsection{Reliability analysis of capacity approximation}
Since power iteration is more susceptible to small perturbations of polynomial coefficients~\cite{wilkinson1963some} than QR-based method, we consider to verify the reliability of our proposed capacity approximation through some experiments.

First of all, we prove that the capacity obtained by our approximation method is consistent with some well-define digraph (that is, $k$-regular graph).
It is easy to calculate the capacity of $k$-regular graphs based on standard definition.
Here, we define 4 graphs of length 2:
\begin{itemize}
\item $\log_2(1)$: a directed cycle, containing ``AC'', ``CG'', ``GT'', and ``TA''.
\item $\log_2(2)$: a GC-balanced digraph, screening ``AA'', ``AT'', ``CC'', ``CG'', ``GC'', ``GG'', ``TA'', and ``TT''.
\item $\log_2(3)$: a digraph without homopolymer, screening ``AA'', ``CC'', ``GG'', and ``TT''.
\item $\log_2(4)$: a complete digraph without screening (4-ary de Bruijn graph of order 2).
\end{itemize}
As the initial results, there are no difference between results approximated by our proposed method and those by oral arithmetic.
And the capacity can be obtained with at most 2 iterations, in other words, the correct result is calculated in the first iteration.

Afterwards, on the basis of the above-mentioned complete graph, we investigate 100 pruned (at random) digraphs for verifying the reliability of our method.
In the automated testing, referential capacities can be approximated by Numpy ``linalg.eig'' function~\cite{anderson1999lapack,van2011numpy}.
The difference between these referential capacities and corresponding approximated capacities can also be used to evaluate the latter's reliability directly.
As shown in Figure~\ref{fig:capacity_evaluation}a and b, 50\% relative error result is between $1.30 \times 10^{-11}$ and $8.16 \times 10^{-11}$.
Besides, the median value of relative error is $2.97 \times 10^{-11}$.
An obvious trend is that the greater the potential information density of a directed graph, the smaller the relative error (Pearson $-0.67$).
For the approximated capacity greater than or equal to 1, the range of relative error is $(2.54 \times 10^{-13}, 2.02 \times 10^{-11})$, which is less than the error tolerance ($10^{-10}$).

Finally, on the basis of the above experiment, we investigate the influence of matrix size change on the relative error.
Here, the only adjustable parameter is that the size of investigated adjacency matrices is gradually increased from $4^2 \times 4^2$ to $4^6 \times 4^6$.
As shown in Figure~\ref{fig:capacity_evaluation}c, no matter how the observed length changes, the median error remains around $3.05 \times 10^{-11}$.
Thus, there is no clear relationship between observed lengths and  errors.
In addition, since each sub-experiment only completed the capacity error comparison of 100 random digraphs, the results cannot be taken as evidence that increasing the observed length can reduce the sudden high error. 

Through these experiments, we believe the relative error of capacity approximation is on the order of minus ten of ten, which is equivalent to error tolerance in the preset values.

\clearpage
\section{Detailed pseudo code for the proposed algorithms in this work}
\label{sec:pseudo_code}

\begin{algorithm}[H]
\caption{Initializing step to build the digraph of the coding algorithm}
\label{al:generating}
\KwIn{de Bruijn graph $\mathcal{D}^k_4$ and constraint set $\mathbb{C}$.}
\KwOut{digraph $\mathcal{D}$ satisfies considered constraints $\mathbb{C}$ and the out-degree requirement.}
Set $\mathcal{D}^k_{\mathbb{C}}$ through deleting all the vertices from $\mathcal{D}^k_4$ which violate one or more constraints from $\mathbb{C}$.\\
Set $\mathcal{D}$ as a digraph induced by the vertex set of $\mathcal{D}^k_{\mathbb{C}}$. \\
Check the out-degree of each vertex in $\mathcal{D}$. \\
If all vertices have out-degree at least 2, jump to step 6; otherwise, go to step 5. \\
Delete all the vertices with out-degree less than 2, then jump to step 3. \\
Output $\mathcal{D}$.
\end{algorithm} 

\begin{algorithm}[H]
\caption{Encoding binary message through graph-based coding algorithm}
\label{al:encoding}
\KwIn{binary message $\bm{x}$.}
\KwOut{DNA sequence $\bm{y}$ satisfying considered constraints $\mathbb{C}$.}
Convert $\bm{x}$ into the decimal number $\pi$.\\
Pick a working vertex in $\mathcal{D}$ virtually~\cite{goldman2013towards,ping2022towards} and set $\bm y$ as an empty sequence. \\
If $\pi$ is greater than $0$, go to step 4; otherwise, jump to step 8. \\
Set $p$ to be the out-degree of the working vertex. \\
Divide $\pi$ by $p$ and let the remainder be $r$ and the quotient be $\lfloor \pi / p \rfloor$. \\
Find the $\{\ast|r\}$-arc and add $\ast\in\{\text{A,C,G,T}\}$ to the end of $\bm{y}$.\\
Set $\pi$ as $\lfloor \pi / p \rfloor$ and set the working vertex as the terminal of the $\{\ast|r\}$-arc, then jump to step 3. \\
Output $\bm{y}$.
\end{algorithm}

\begin{algorithm}[H]
\caption{Repair DNA sequence through path-based error correcting}
\label{al:repairing}
\KwIn{wrong DNA sequence $\bm{y}_\text{wrong}$, path check sequence $\bm{y}_\text{check}$, and digraph $\mathcal{D}$.}
\KwOut{repaired DNA sequence set $\bm{Y}_\text{repair}$.}
Create two empty set $\bm{Y}_\text{search}$ and $\bm{Y}_\text{repair}$. \\
Put $\bm{y}_\text{wrong}$ into an empty candidate set $\bm{Y}_\text{candidate}$. \\
If $\bm{Y}_\text{candidate}$ becomes an empty set, jump to step 9; otherwise, go to step 4. \\
Take out a candidate DNA sequence $\bm{y}_\text{candidate}$ from $\bm{Y}_\text{candidate}$. \\
If $\bm{y}_\text{candidate}$ belongs to a path in $\mathcal{D}$, put $\bm{y}_\text{candidate}$ into $\bm{Y}_\text{search}$ and jump to step 3; otherwise, go to step 6. \\
Obtain $p$ as the position where the path error first occurs in $\bm{y}_\text{candidate}$. \\
Use $\mathcal{D}$ to do local exhaustive reverse search for the position $p$ to position $(p - k)$ of $\bm{y}_\text{candidate}$. \\
Collect all local repairable DNA sequences in step 7 into $\bm{Y}_\text{candidate}$, then jump to step 3. \\
Put DNA sequences that their path check sequence equals $\bm{y}_\text{check}$ from $\bm{Y}_\text{search}$ into $\bm{Y}_\text{repair}$. \\
Output $\bm{Y}_\text{repair}$.
\end{algorithm}

\begin{algorithm}[H]
\caption{Approximate the capacity}
\label{al:approximation}
\KwIn{accessor $\alpha$.}
\KwOut{approximated capacity $\rho$.}
Set $i$ to be $0$ and set the initial eigenvector $\gamma_0$ as the all-one vector $(1,1, \cdots, 1,1)_{1 \times 4^k}$. \\
Set $j$ to be $1$ and set $\gamma_{i + 1}$ as the all-zero vector $(0,0, \cdots, 0,0)_{1 \times 4^k}$. \\ 
Let $\gamma_k$ as the key vector contains indices of the its value in $\alpha^\text{T}[j]$ greater than 0. \\
Let $\gamma_v$ as the value vector contains values for location $\gamma_k$ in $\gamma_i$. \\ 
Set $\gamma_{i + 1}[\gamma_k]$ as $\gamma_{i + 1}[\gamma_k] + \gamma_v$ and set $j$ as $j + 1$. \\
If $j$ is $4$, go to step 7; otherwise, go to step 3. \\
If $\frac{|\max(\gamma_{i + 1}) - \max(\gamma_i)|}{\max(\gamma_i)} \leq 10^{-10}$, go to step 8; otherwise, set $i$ as $i + 1$ and go to step 2. \\
Output $\log_2 \{\max (\gamma_{i + 1})\}$.
\end{algorithm}

\clearpage

\section{Supporting figures and tables}
\label{sec:addition_information}

\begin{table}[htbp]
\centering
\caption{\textbf{Representative regional biochemical constraint sets.} 
There are 12 sets of constraints sorted by the capacity, the approximation of which is mentioned in Appendix~\ref{sec:math_proof}. 
$^\ast$ N/A represents the constraint set does not contain this type of constraint. }  
\label{tab:constraint_sets}
\begin{tabular}{lllll}
\hline
\textbf{set index} & \textbf{homopolymer run-length} & \textbf{regionalized GC content} & \textbf{undesired motifs} & \textbf{capacity} \\
\hline
01 & 2 & 50\% & restriction enzyme sites & 1.0000 \\
02 & 1 & N/A$^\ast$ & N/A & 1.5850 \\
03 & N/A & 10\% -- 30\%$^\dag$ & N/A & 1.6302 \\
04 & 2 & 40\% -- 60\% & high error motifs in ONT sequencer & 1.6698  \\
05 & 2 & 40\% -- 60\% & N/A & 1.7761 \\
06 & N/A & 50\% -- 70\%$^\dag$ & N/A & 1.7958 \\
07 & 3 & 40\% -- 60\% & N/A & 1.8114 \\
08 & 4 & 40\% -- 60\% & N/A & 1.8152 \\
09 & 3 & N/A & N/A & 1.9824 \\
10 & 4 & N/A & N/A & 1.9957\\
11 & 5 & N/A & N/A & 1.9989\\
12 & 6 & N/A & N/A & 1.9997\\
\hline
\end{tabular}
\end{table}

\clearpage

\clearpage

\begin{figure}[ht]
    \begin{center}
    \includegraphics[width=1\columnwidth]{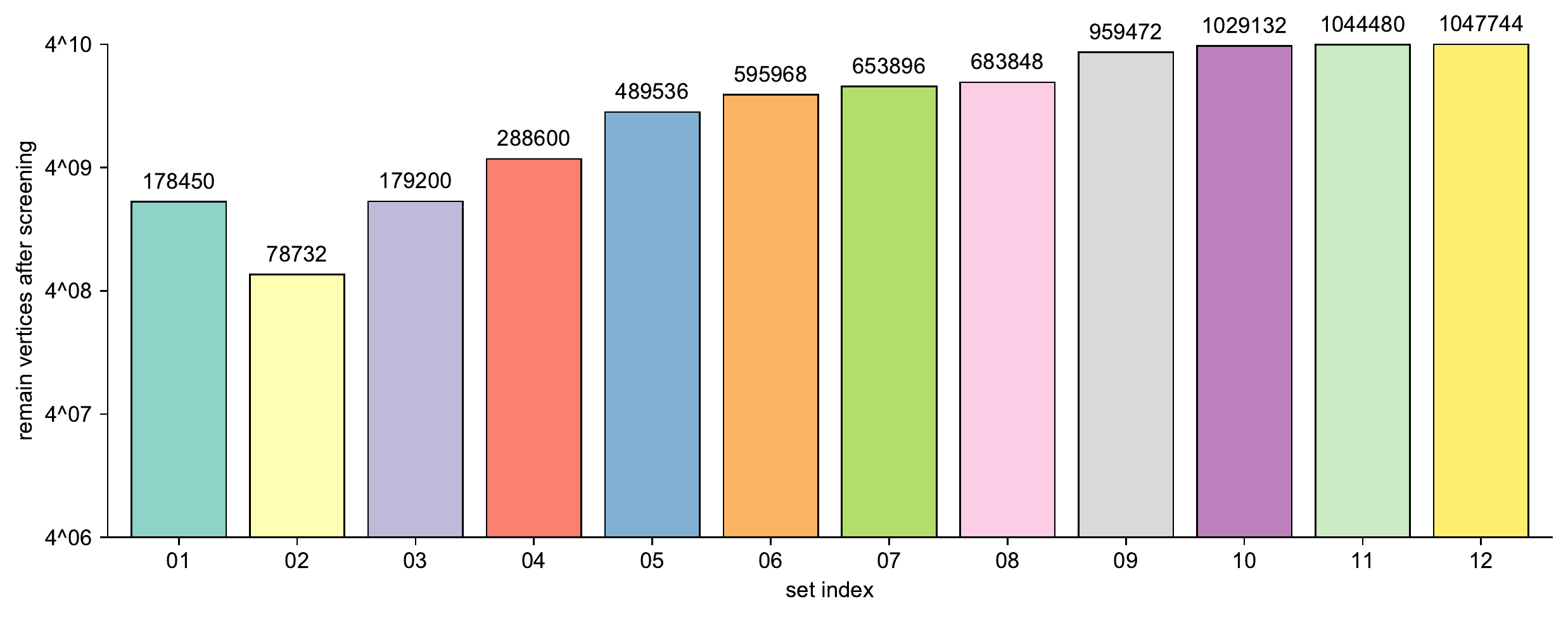}
    \caption{\textbf{Remaining vertex number after screening process of SPIDER-WEB.}
    In the generation process, the screening step is currently accounts for 80\% or more of the time.}
    \label{fig:screening_vertex}
    \end{center}
\end{figure}

\clearpage

\begin{landscape}
\begin{table}[htbp]
\centering
\caption{\textbf{Remaining vertex number during trimming process of SPIDER-WEB.} 
$^\ast$ N/A represents the trimming process is finished. }  
\label{tab:trimming_vertex}
\begin{tabular}{lllllllllllll}
\hline
\textbf{parameter index} & \multicolumn{12}{l}{\textbf{constraint set}} \\
& 01 & 02 & 03 & 04 & 05 & 06 & 07 & 08 & 09 & 10 & 11 & 12 \\
\hline
0 & 162278 & 78732 & 179200 & 279248 & 486032 & 595968 & 653512 & 683848 & 959472 & 1029132 & 1044480 & 1047744 \\
1 & 132626 & N/A$^\ast$ & N/A & 274144 & 479232 & N/A & 652744 & N/A & N/A & N/A & N/A & N/A \\
2 & 106131 & N/A & N/A & 270672 & 474392 & N/A & 651208 & N/A & N/A & N/A & N/A & N/A \\
3 & 74843 & N/A & N/A & 270232 & N/A & N/A & 649864 & N/A & N/A & N/A & N/A & N/A \\
4 & 52508 & N/A & N/A & N/A & N/A & N/A & N/A & N/A & N/A & N/A & N/A & N/A \\
5 & 33878 & N/A & N/A & N/A & N/A & N/A & N/A & N/A & N/A & N/A & N/A & N/A \\
6 & 22632 & N/A & N/A & N/A & N/A & N/A & N/A & N/A & N/A & N/A & N/A & N/A \\
7 & 14968 & N/A & N/A & N/A & N/A & N/A & N/A & N/A & N/A & N/A & N/A & N/A \\
8 & 10048 & N/A & N/A & N/A & N/A & N/A & N/A & N/A & N/A & N/A & N/A & N/A \\
9 & 6912 & N/A & N/A & N/A & N/A & N/A & N/A & N/A & N/A & N/A & N/A & N/A\\
10 & 4608 & N/A & N/A & N/A & N/A & N/A & N/A & N/A & N/A & N/A & N/A & N/A \\
11 & 3072 & N/A & N/A & N/A & N/A & N/A & N/A & N/A & N/A & N/A & N/A & N/A\\
12 & 2048 & N/A & N/A & N/A & N/A & N/A & N/A & N/A & N/A & N/A & N/A & N/A\\
\hline
\end{tabular}
\end{table}
\end{landscape}

\clearpage

\begin{center}
\begin{longtable}{l|ll|l|ll}
\caption{\textbf{Adjustable parameter list of well-established coding algorithms in the coding task.} 
Remaining parameters in these algorithms use their default values.
$^\ast$ $c$ and $\delta$ in Soliton distribution~\cite{luby2002lt}, are the main factors in DNA Fountain. Their interval~\cite{hyytia2006optimizing} are set as $[0.1, 1.0]$ and $[0.01, 0.1]$ respectively.
$^\dag$ Yin-Yang Code contains 6,144 rules~\cite{ping2022towards}.
$^\ddag$ HEDGES provides 6 repetitive patterns and their corresponding correct penalty~\cite{press2020hedges}.} 
\label{tab:parameter_1}
\\ \hline
\textbf{parameter index} & \multicolumn{2}{l|}{\textbf{DNA fountain}$^\ast$} & \textbf{Yin-Yang Code}$^\dag$ & \multicolumn{2}{|l}{\textbf{HEDGES}$^\ddag$}\\
 & $c$ & $\delta$ & rule index & pattern index & correct penalty \\
\hline
\endfirsthead

\hline
\textbf{parameter index} & \multicolumn{2}{l|}{\textbf{DNA fountain}} & \textbf{Yin-Yang Code} & \multicolumn{2}{|l}{\textbf{HEDGES}} \\
 & $c$ & $\delta$ & rule index & pattern index & correct penalty \\
\hline
\endhead

\hline 
\multicolumn{6}{r}{Continued on next page} \\ 
\hline
\endfoot

\hline
\endlastfoot
1 & 0.65 & 0.08 & 3322 & $-[2-1]-$ & -0.035 \\
2 & 0.23 & 0.04 & 2223 & $-[2-1-1-1-1]-$ & -0.082 \\
3 & 1.00 & 0.02 & 4501 & $-[1]-$ & -0.127 \\
4 & 0.26 & 0.08 & 4815 & $-[1-1-0]-$ & -0.229 \\
5 & 0.70 & 0.08 & 676 & $-[1-0]-$ & -0.265 \\
6 & 0.19 & 0.02 & 3796 & $-[1-0-0]-$ & -0.324 \\
7 & 0.97 & 0.07 & 1548 &   &  \\
8 & 0.18 & 0.06 & 3281 &   &  \\
9 & 0.65 & 0.10 & 2145 &   &  \\
10 & 0.62 & 0.04 & 3602 &   &  \\
11 & 0.51 & 0.03 & 4186 &   &  \\
12 & 0.61 & 0.03 & 1731 &   &  \\
13 & 0.63 & 0.05 & 504 &   &  \\
14 & 0.57 & 0.08 & 3173 &   &  \\
15 & 0.76 & 0.02 & 5148 &   &  \\
16 & 0.70 & 0.07 & 53 &   &  \\
17 & 0.85 & 0.03 & 285 &   &  \\
18 & 0.66 & 0.07 & 2467 &   &  \\
19 & 0.37 & 0.09 & 1740 &   &  \\
20 & 0.63 & 0.10 & 3122 &   &  \\
21 & 0.50 & 0.02 & 1579 &   &  \\
22 & 0.56 & 0.05 & 5400 &   &  \\
23 & 0.92 & 0.07 & 4705 &   &  \\
24 & 0.90 & 0.06 & 650 &   &  \\
25 & 0.67 & 0.04 & 1638 &   &  \\
26 & 0.84 & 0.04 & 4755 &   &  \\
27 & 0.38 & 0.03 & 4079 &   &  \\
28 & 0.34 & 0.07 & 4706 &   &  \\
29 & 0.24 & 0.04 & 2039 &   &  \\
30 & 0.68 & 0.09 & 1182 &   &  \\
31 & 0.16 & 0.05 & 32 &   &  \\
32 & 0.40 & 0.07 & 3398 &   &  \\
33 & 0.53 & 0.08 & 5986 &   &  \\
34 & 0.18 & 0.09 & 2337 &   &  \\
35 & 0.80 & 0.08 & 3046 &   &  \\
36 & 0.16 & 0.06 & 4332 &   &  \\
37 & 0.81 & 0.03 & 1188 &   &  \\
38 & 0.43 & 0.09 & 2155 &   &  \\
39 & 0.71 & 0.08 & 2758 &   &  \\
40 & 0.48 & 0.04 & 2197 &   &  \\
41 & 0.44 & 0.03 & 2046 &   &  \\
42 & 0.71 & 0.06 & 5298 &   &  \\
43 & 0.97 & 0.05 & 986 &   &  \\
44 & 0.40 & 0.10 & 3995 &   &  \\
45 & 0.84 & 0.04 & 636 &   &  \\
46 & 0.84 & 0.06 & 1040 &   &  \\
47 & 0.75 & 0.09 & 3900 &   &  \\
48 & 0.55 & 0.06 & 3869 &   &  \\
49 & 0.68 & 0.04 & 2505 &   &  \\
50 & 0.66 & 0.05 & 898 &   &  \\
51 & 0.21 & 0.10 & 6040 &   &  \\
52 & 0.51 & 0.02 & 4851 &   &  \\
53 & 0.84 & 0.03 & 3547 &   &  \\
54 & 0.31 & 0.07 & 773 &   &  \\
55 & 0.75 & 0.02 & 3219 &   &  \\
56 & 0.45 & 0.08 & 5112 &   &  \\
57 & 0.83 & 0.08 & 253 &   &  \\
58 & 0.87 & 0.04 & 6049 &   &  \\
59 & 0.24 & 0.07 & 4432 &   &  \\
60 & 0.76 & 0.02 & 5030 &   &  \\
61 & 0.34 & 0.04 & 4029 &   &  \\
62 & 0.55 & 0.07 & 1810 &   &  \\
63 & 0.78 & 0.02 & 4860 &   &  \\
64 & 0.29 & 0.07 & 5775 &   &  \\
65 & 0.91 & 0.08 & 236 &   &  \\
66 & 0.83 & 0.06 & 1511 &   &  \\
67 & 0.63 & 0.09 & 2243 &   &  \\
68 & 0.67 & 0.09 & 2841 &   &  \\
69 & 0.47 & 0.07 & 5275 &   &  \\
70 & 0.91 & 0.02 & 698 &   &  \\
71 & 0.71 & 0.02 & 4103 &   &  \\
72 & 0.29 & 0.09 & 144 &   &  \\
73 & 0.57 & 0.07 & 1886 &   &  \\
74 & 0.88 & 0.08 & 904 &   &  \\
75 & 0.71 & 0.06 & 5645 &   &  \\
76 & 0.82 & 0.01 & 3383 &   &  \\
77 & 0.47 & 0.10 & 2392 &   &  \\
78 & 0.95 & 0.07 & 944 &   &  \\
79 & 0.76 & 0.02 & 1236 &   &  \\
80 & 0.65 & 0.02 & 2796 &   &  \\
81 & 0.64 & 0.05 & 3203 &   &  \\
82 & 0.48 & 0.09 & 3079 &   &  \\
83 & 0.79 & 0.01 & 2886 &   &  \\
84 & 0.77 & 0.09 & 2797 &   &  \\
85 & 0.95 & 0.02 & 887 &   &  \\
86 & 0.69 & 0.04 & 98 &   &  \\
87 & 0.48 & 0.06 & 3969 &   &  \\
88 & 0.65 & 0.05 & 1425 &   &  \\
89 & 0.37 & 0.07 & 1509 &   &  \\
90 & 0.63 & 0.03 & 1288 &   &  \\
91 & 0.54 & 0.03 & 4480 &   &  \\
92 & 0.99 & 0.03 & 6035 &   &  \\
93 & 0.81 & 0.05 & 5110 &   &  \\
94 & 0.73 & 0.06 & 3408 &   &  \\
95 & 0.76 & 0.08 & 2001 &   &  \\
96 & 0.36 & 0.04 & 3640 &   &  \\
97 & 0.10 & 0.05 & 4902 &   &  \\
98 & 0.90 & 0.08 & 1554 &   &  \\
99 & 0.40 & 0.02 & 1688 &   &  \\
100 & 0.34 & 0.07 & 2914 &   &  \\
\end{longtable}
\end{center}

\clearpage

\begin{landscape}
\begin{center}
\begin{longtable}{l|llllllllllll}
\caption{\textbf{Adjustable parameter list of SPIDER-WEB in the coding task.} 
Since the difference of virtual vertices may affect the performance of generated coding algorithms, we adjust the virtual vertices 100 times here.
Under different constraint sets, the number of vertices contained in the digraph of the coding algorithm is different, so the virtual vertex index corresponding to the parameter index will be inconsistent.
} 
\label{tab:parameter_2}
\\ \hline
\textbf{parameter index} & \multicolumn{12}{l}{\textbf{constraint set}} \\
& 01 & 02 & 03 & 04 & 05 & 06 & 07 & 08 & 09 & 10 & 11 & 12 \\
\hline
\endfirsthead

\hline
\textbf{parameter index} & \multicolumn{12}{l}{\textbf{constraint set}} \\
& 01 & 02 & 03 & 04 & 05 & 06 & 07 & 08 & 09 & 10 & 11 & 12 \\
\hline
\endhead

\hline 
\multicolumn{13}{r}{Continued on next page} \\ 
\hline
\endfoot

\hline
\endlastfoot

1 & 142829 & 866161 & 253044 & 281093 & 178452 & 218484 & 89731 & 904827 & 684145 & 687192 & 124899 & 885640 \\
2 & 187682 & 580716 & 29007 & 165608 & 469807 & 965801 & 845153 & 958909 & 904444 & 882829 & 120275 & 264391 \\
3 & 863710 & 912818 & 831488 & 675005 & 188845 & 218640 & 315842 & 659002 & 1022437 & 1014826 & 802316 & 707094 \\
4 & 187694 & 160845 & 204668 & 329351 & 219616 & 283332 & 827022 & 458234 & 478346 & 433611 & 320547 & 949802 \\
5 & 971502 & 161950 & 262973 & 776559 & 1002896 & 409813 & 22208 & 250506 & 140772 & 781445 & 437723 & 839263 \\
6 & 559947 & 845006 & 12831 & 71951 & 412638 & 26885 & 858572 & 144494 & 212722 & 663479 & 255500 & 1025438 \\
7 & 74205 & 756509 & 197103 & 529626 & 904262 & 502770 & 330678 & 810855 & 54512 & 521629 & 76517 & 232697 \\
8 & 184594 & 624793 & 455884 & 48058 & 187878 & 482143 & 1018721 & 927817 & 30332 & 893363 & 470154 & 709828 \\
9 & 756551 & 570514 & 81910 & 299307 & 390395 & 993952 & 914678 & 899887 & 1026096 & 567603 & 956118 & 704079 \\
10 & 188113 & 400860 & 996736 & 592720 & 728009 & 641528 & 382092 & 670411 & 330298 & 239518 & 116328 & 204835 \\
11 & 555908 & 844268 & 593459 & 823604 & 42221 & 873346 & 900814 & 527041 & 903544 & 151315 & 450349 & 754247 \\
12 & 554167 & 629917 & 975155 & 222990 & 465649 & 435265 & 226842 & 413052 & 859213 & 930436 & 65612 & 1045379 \\
13 & 922093 & 812263 & 945407 & 318262 & 482461 & 1025893 & 456381 & 135805 & 207805 & 107109 & 444059 & 631952 \\
14 & 489287 & 280262 & 208524 & 44794 & 396933 & 55787 & 86811 & 524773 & 112056 & 253689 & 571129 & 915255 \\
15 & 767156 & 924537 & 954168 & 925582 & 1000866 & 158635 & 424003 & 266993 & 47255 & 1022416 & 842915 & 1029938 \\
16 & 135714 & 585188 & 313137 & 809523 & 231118 & 650720 & 938631 & 859866 & 75107 & 511942 & 8104 & 751048 \\
17 & 974382 & 298614 & 524089 & 682960 & 195466 & 735461 & 83342 & 363692 & 281639 & 841301 & 638159 & 882168 \\
18 & 571572 & 211428 & 250060 & 763420 & 734952 & 841094 & 347687 & 1027291 & 332363 & 766314 & 504047 & 910443 \\
19 & 975134 & 629879 & 60929 & 823384 & 769129 & 546925 & 770679 & 869567 & 381459 & 953287 & 578507 & 969848 \\
20 & 123617 & 308155 & 67312 & 533712 & 770990 & 362050 & 795653 & 412808 & 263675 & 121381 & 483222 & 77009 \\
21 & 73453 & 201432 & 64761 & 149836 & 997642 & 749213 & 171226 & 108623 & 944188 & 750295 & 148877 & 304640 \\
22 & 971041 & 299852 & 134973 & 413761 & 234552 & 368349 & 320482 & 593118 & 123527 & 163608 & 568638 & 748596 \\
23 & 859694 & 466508 & 800718 & 798025 & 17699 & 940578 & 225451 & 638334 & 112794 & 545121 & 704711 & 839136 \\
24 & 119249 & 405108 & 1030139 & 330694 & 91683 & 31640 & 150007 & 750152 & 296518 & 728116 & 351268 & 836158 \\
25 & 70369 & 922547 & 540417 & 994122 & 853629 & 832877 & 934006 & 561400 & 313357 & 937376 & 81771 & 781148 \\
26 & 975581 & 191604 & 230272 & 235068 & 318530 & 538014 & 971281 & 192870 & 239970 & 460583 & 1036163 & 628843 \\
27 & 929070 & 559907 & 602300 & 97187 & 470770 & 712938 & 815204 & 594716 & 231384 & 747543 & 125729 & 553667 \\
28 & 909010 & 929043 & 8312 & 191271 & 153777 & 373784 & 222450 & 876696 & 426820 & 758262 & 717273 & 1000814 \\
29 & 119085 & 191587 & 864515 & 757435 & 274365 & 347876 & 857925 & 925315 & 727588 & 348894 & 622064 & 760074 \\
30 & 739464 & 640948 & 102607 & 333343 & 648250 & 1028517 & 992609 & 505404 & 704155 & 339916 & 972777 & 165433 \\
31 & 281412 & 231186 & 246978 & 657674 & 656484 & 420483 & 466285 & 54134 & 421790 & 86992 & 570664 & 855486 \\
32 & 860641 & 321262 & 20015 & 709551 & 592807 & 87315 & 165074 & 165389 & 778889 & 469997 & 600202 & 532651 \\
33 & 77294 & 121741 & 816190 & 380275 & 951504 & 530033 & 867061 & 514488 & 548197 & 813443 & 577537 & 966618 \\
34 & 571320 & 204505 & 983158 & 768723 & 334772 & 645767 & 625743 & 192474 & 470394 & 1040474 & 454446 & 1046294 \\
35 & 542855 & 847134 & 8152 & 741626 & 899407 & 784650 & 650334 & 936669 & 113178 & 1001491 & 41222 & 895118 \\
36 & 188705 & 319028 & 65660 & 117031 & 138973 & 775931 & 296715 & 747599 & 664941 & 909551 & 553551 & 1030002 \\
37 & 756619 & 75704 & 968718 & 748359 & 180624 & 36906 & 653963 & 253796 & 1031542 & 564075 & 435918 & 653677 \\
38 & 494472 & 945443 & 527431 & 410407 & 743478 & 808366 & 194542 & 213389 & 562017 & 346057 & 378550 & 807740 \\
39 & 296824 & 186802 & 880703 & 114284 & 869449 & 71124 & 755859 & 460250 & 204536 & 496999 & 1010384 & 958988 \\
40 & 296075 & 242467 & 980162 & 380336 & 463654 & 388282 & 657870 & 165370 & 1009945 & 1014545 & 997186 & 487107 \\
41 & 908573 & 506151 & 196087 & 341437 & 339432 & 428177 & 979298 & 75231 & 316393 & 966421 & 103218 & 231209 \\
42 & 971489 & 937250 & 12407 & 399857 & 560541 & 178519 & 641900 & 522125 & 517359 & 21370 & 823458 & 163350 \\
43 & 308360 & 604958 & 226300 & 214468 & 222673 & 283676 & 658122 & 146886 & 288116 & 573375 & 580914 & 227544 \\
44 & 908830 & 412793 & 668595 & 834972 & 215005 & 346220 & 678990 & 924136 & 193388 & 763785 & 159495 & 870084 \\
45 & 909789 & 142882 & 984284 & 975262 & 757543 & 359084 & 269852 & 234528 & 149618 & 358748 & 90144 & 889457 \\
46 & 279735 & 596088 & 581596 & 22608 & 988308 & 88616 & 292574 & 280023 & 500606 & 35596 & 351410 & 832300 \\
47 & 555912 & 243245 & 396344 & 468355 & 883645 & 433618 & 957939 & 613625 & 193172 & 67084 & 568515 & 597785 \\
48 & 489652 & 116536 & 12748 & 630520 & 535647 & 239286 & 187779 & 684255 & 69709 & 353076 & 247629 & 651936 \\
49 & 926237 & 203063 & 995155 & 129112 & 1024741 & 418012 & 913492 & 901573 & 953183 & 278713 & 312755 & 861967 \\
50 & 297092 & 767841 & 206543 & 514463 & 405324 & 304351 & 1025003 & 206457 & 802630 & 89671 & 350094 & 944446 \\
51 & 123614 & 116663 & 837661 & 509467 & 35957 & 882283 & 648186 & 714611 & 566226 & 146055 & 188240 & 316636 \\
52 & 978205 & 603748 & 134907 & 496407 & 516231 & 161944 & 568896 & 924044 & 869920 & 543654 & 773776 & 382897 \\
53 & 922157 & 190324 & 795487 & 988768 & 189371 & 129895 & 708236 & 149544 & 312352 & 616954 & 577367 & 655213 \\
54 & 493687 & 506669 & 822716 & 153807 & 326635 & 180693 & 374583 & 604086 & 608875 & 88901 & 179253 & 628669 \\
55 & 490315 & 214940 & 799289 & 37928 & 336672 & 941991 & 135970 & 1014404 & 931390 & 547346 & 509187 & 1039606 \\
56 & 476228 & 549153 & 12012 & 730258 & 286896 & 144937 & 570638 & 1011270 & 482463 & 560916 & 760321 & 1046029 \\
57 & 905773 & 214833 & 294867 & 343315 & 84423 & 281023 & 465644 & 979486 & 727619 & 708803 & 358467 & 447680 \\
58 & 559943 & 859704 & 848203 & 398133 & 240912 & 215694 & 301090 & 4512 & 846138 & 855888 & 313921 & 937641 \\
59 & 490360 & 111751 & 838112 & 211897 & 444507 & 405483 & 609714 & 315882 & 928383 & 554432 & 771451 & 719616 \\
60 & 908753 & 637875 & 721983 & 159457 & 226420 & 762395 & 470730 & 309716 & 400965 & 148903 & 472477 & 903030 \\
61 & 143058 & 766343 & 255212 & 757950 & 823418 & 225194 & 65209 & 730256 & 30277 & 463128 & 294539 & 883861 \\
62 & 912925 & 402023 & 795155 & 457017 & 921153 & 281689 & 742360 & 447729 & 1012559 & 12394 & 226800 & 631668 \\
63 & 184797 & 472612 & 111407 & 847386 & 710588 & 157056 & 338372 & 590672 & 576895 & 857181 & 698151 & 795196 \\
64 & 122349 & 805671 & 574 & 417875 & 947230 & 362930 & 297477 & 828999 & 163470 & 171858 & 78001 & 456876 \\
65 & 476279 & 214855 & 950771 & 846829 & 289741 & 501987 & 826106 & 504728 & 384157 & 175664 & 126489 & 356784 \\
66 & 856541 & 318665 & 985330 & 1000359 & 530268 & 935863 & 67092 & 539581 & 793407 & 813090 & 802192 & 253022 \\
67 & 740168 & 563319 & 525192 & 214658 & 869495 & 346901 & 930148 & 578404 & 848509 & 572612 & 153821 & 128441 \\
68 & 768072 & 73116 & 210759 & 77108 & 388127 & 106781 & 925975 & 660558 & 70905 & 548894 & 324767 & 830820 \\
69 & 905949 & 486499 & 820043 & 817974 & 604677 & 615059 & 401614 & 792569 & 503091 & 683161 & 571958 & 904335 \\
70 & 859873 & 549747 & 209666 & 603443 & 181995 & 422481 & 667832 & 834954 & 102215 & 515048 & 270008 & 164505 \\
71 & 143074 & 421422 & 247788 & 179525 & 466865 & 669828 & 996446 & 30888 & 556696 & 514246 & 813100 & 1024614 \\
72 & 122593 & 907705 & 532499 & 115096 & 713805 & 230725 & 1015074 & 781982 & 750093 & 146630 & 508809 & 247976 \\
73 & 143073 & 889708 & 811855 & 638470 & 765566 & 751802 & 55935 & 585631 & 126320 & 800754 & 155607 & 833891 \\
74 & 123410 & 290424 & 64555 & 1016670 & 408023 & 21679 & 925210 & 500862 & 499833 & 794879 & 1013685 & 719196 \\
75 & 856365 & 112161 & 211963 & 968626 & 472999 & 1022932 & 649931 & 998747 & 696272 & 42417 & 593384 & 417506 \\
76 & 913117 & 411421 & 196923 & 723406 & 20569 & 960456 & 122516 & 59574 & 718980 & 861612 & 198900 & 398299 \\
77 & 122577 & 478918 & 36094 & 500787 & 1000544 & 304385 & 925589 & 933252 & 533617 & 182774 & 591584 & 690235 \\
78 & 476344 & 142968 & 983108 & 609546 & 935803 & 993944 & 75928 & 218374 & 345630 & 128512 & 374396 & 342524 \\
79 & 767111 & 728475 & 1044575 & 128693 & 865867 & 686449 & 316149 & 99070 & 643690 & 654034 & 1029719 & 476325 \\
80 & 554887 & 406316 & 838396 & 375033 & 133988 & 273090 & 946663 & 581259 & 994514 & 871352 & 868740 & 89782 \\
81 & 477060 & 580071 & 200567 & 945735 & 1009622 & 885163 & 880753 & 900073 & 21180 & 174973 & 106770 & 415749 \\
82 & 70110 & 908365 & 12228 & 615484 & 775122 & 365939 & 150014 & 159853 & 632616 & 26604 & 876508 & 761111 \\
83 & 768948 & 291619 & 793712 & 844697 & 282930 & 92784 & 230797 & 101662 & 506338 & 1012454 & 173458 & 911376 \\
84 & 751755 & 470747 & 970892 & 168390 & 144924 & 243190 & 815081 & 583378 & 705398 & 85595 & 990231 & 1001526 \\
85 & 859437 & 419037 & 56547 & 731375 & 925620 & 1022816 & 118506 & 954404 & 121934 & 773538 & 855814 & 345373 \\
86 & 863709 & 582503 & 2604 & 519814 & 888105 & 89725 & 154963 & 495065 & 95836 & 804894 & 646714 & 4438 \\
87 & 860897 & 549170 & 47027 & 158676 & 749334 & 148507 & 944692 & 791684 & 586819 & 609992 & 198431 & 580374 \\
88 & 978478 & 473186 & 181440 & 776544 & 1015329 & 506484 & 1013739 & 486288 & 375772 & 634539 & 743352 & 550842 \\
89 & 135441 & 805809 & 196564 & 717998 & 496560 & 407979 & 330215 & 964046 & 681193 & 214089 & 34613 & 848329 \\
90 & 859426 & 203038 & 259292 & 718495 & 397446 & 858411 & 61039 & 184683 & 739265 & 521790 & 895788 & 527861 \\
91 & 974110 & 298062 & 847612 & 1025211 & 208353 & 570838 & 92252 & 80186 & 346899 & 424906 & 859156 & 752102 \\
92 & 489396 & 136121 & 315074 & 397164 & 244453 & 773016 & 570397 & 331667 & 402157 & 908542 & 947640 & 433751 \\
93 & 856354 & 187928 & 573390 & 812826 & 899517 & 716472 & 569937 & 319997 & 527782 & 484916 & 94450 & 346352 \\
94 & 542907 & 472940 & 604 & 951396 & 50461 & 148311 & 195975 & 709584 & 301591 & 801060 & 918823 & 94075 \\
95 & 135918 & 444642 & 1044880 & 923279 & 364728 & 385333 & 537366 & 132690 & 279388 & 912853 & 678735 & 89278 \\
96 & 493428 & 496364 & 48137 & 578642 & 33034 & 372411 & 325742 & 396600 & 832929 & 491478 & 890077 & 57363 \\
97 & 908578 & 145292 & 930612 & 104910 & 1024998 & 702786 & 804710 & 114654 & 997019 & 393918 & 266400 & 658449 \\
98 & 505784 & 585138 & 261649 & 375930 & 380391 & 751442 & 7005 & 979278 & 868565 & 838733 & 449503 & 836757 \\
99 & 860690 & 477049 & 198575 & 953956 & 580158 & 500862 & 342056 & 1000609 & 390934 & 807357 & 709216 & 188477 \\
100 & 297035 & 814393 & 999027 & 455041 & 416739 & 18985 & 396114 & 493048 & 49590 & 473233 & 241415 & 14907 \\
\end{longtable}
\end{center}
\end{landscape}

\clearpage

\begin{table}[htbp]
\centering
\caption{\textbf{Coding performance of different coding algorithms.} 
N/A means that the algorithm cannot finish the encode process under such constraint set.}  
\label{tab:coding_performance}
\begin{tabular}{lllll}
\hline
\textbf{set index} & \textbf{DNA Fountain} & \textbf{Yin-Yang Code} & \textbf{HEDGES} & \textbf{SPIDER-WEB} \\
\hline
01 & N/A & 0.89 & 0.30 -- 0.89 & 0.92 \\
02 & N/A & 0.89 & 0.30 -- 1.33 & 1.46 \\
03 & N/A & 0.89 & 0.30 -- 1.07 & 1.37 -- 1.40 \\
04 & N/A & 0.89 & 0.30 -- 1.33 & 1.44 -- 1.48  \\
05 & N/A & 0.89 & 0.30 -- 1.33 & 1.56 -- 1.59 \\
06 & N/A & 0.89 & 0.30 -- 1.33 & 1.55 -- 1.59 \\
07 & N/A & 0.89 & 0.30 -- 1.33 & 1.49 -- 1.54 \\
08 & N/A & 0.89 & 0.30 -- 1.33 & 1.58 -- 1.62 \\
09 & 1.06 -- 1.73 & 0.89 -- 1.77 & 0.30 -- 1.33 & 1.78 -- 1.79 \\
10 & 1.40 -- 1.73 & 1.64 -- 1.78 & 0.30 -- 1.33 & 1.79 -- 1.80 \\
11 & 1.45 -- 1.73 & 1.72 -- 1.78 & 0.30 -- 1.33 & 1.79 -- 1.80 \\
12 & 1.47 -- 1.73 & 1.75 -- 1.78 & 0.30 -- 1.33 & 1.81 -- 1.82 \\
\hline
\end{tabular}
\end{table}

\clearpage

\begin{table}[htbp]
\centering
\caption{\textbf{Standard deviation of different coding algorithm performances.} 
$^\ast$ N/A refers to the standard deviation cannot be calculated. 
$^\dag$ The information densities under the successful decoding are selected. }  
\label{tab:stability}
\begin{tabular}{lllll}
\hline
\textbf{set index} & \textbf{DNA Fountain} & \textbf{Yin-Yang Code} & \textbf{HEDGES}$^\dag$ & \textbf{SPIDER-WEB} \\
\hline
01 & N/A$^\ast$ & 0.0000 & 0.1936 & 0.0000 \\
02 & N/A & 0.0000 & 0.2757 & 0.0000 \\
03 & N/A & 0.0000 & 0.2417 & 0.0079 \\
04 & N/A & 0.0000 & 0.2547 & 0.0081 \\
05 & N/A & 0.0000 & 0.2849 & 0.0075 \\
06 & N/A & 0.0000 & 0.2741 & 0.0083 \\
07 & N/A & 0.0000 & 0.2900 & 0.0109 \\
08 & N/A & 0.0000 & 0.2865 & 0.0082 \\
09 & 0.0595 & 0.3896 & 0.3574 & 0.0008 \\
10 & 0.0469 & 0.0525 & 0.3609 & 0.0003 \\
11 & 0.0436 & 0.0229 & 0.3609 & 0.0007 \\
12 & 0.0386 & 0.0110 & 0.3609 & 0.0011 \\
\hline
\end{tabular}
\end{table}

\clearpage

\begin{table}[htbp]
\centering
\caption{\textbf{Decoding success probability of repetitive patterns under the error-free retrieval.} 
In HEDGES, a repetitive pattern shows a cycled array with a given length, each value in the array is the number of bits represented by the current nucleotide.
Pattern 1 to 6 are $-[2-1]-$, $-[2-1-1-1-1]-$, $-[1]-$, $-[1-1-0]-$, $-[1-0]-$, and $-[1-0-0]-$, respectively.}  
\label{tab:hedges_rate}
\begin{tabular}{lllllll}
\hline
\textbf{set index} & \textbf{pattern 1} & \textbf{pattern 2} & \textbf{pattern 3} & \textbf{pattern 4} & \textbf{pattern 5} & \textbf{pattern 6} \\
\hline
01 & 0\% & 0\% & 8\% & 10\% & 11\% & 19\% \\
02 & 5\% & 60\% & 100\% & 100\% & 100\% & 100\% \\
03 & 0\% & 22\% & 100\% & 100\% & 100\% & 100\% \\
04 & 4\% & 22\% & 74\% & 85\% & 86\% & 92\% \\
05 & 14\% & 48\% & 92\% & 96\% & 97\% & 99\% \\
06 & 6\% & 53\% & 100\% & 100\% & 100\% & 100\% \\
07 & 16\% & 54\% & 98\% & 99\% & 100\% & 100\% \\
08 & 14\% & 52\% & 100\% & 100\% & 100\% & 100\% \\
09 & 92\% & 99\% & 100\% & 100\% & 100\% & 100\% \\
10 & 100\% & 100\% & 100\% & 100\% & 100\% & 100\% \\
11 & 100\% & 100\% & 100\% & 100\% & 100\% & 100\% \\
12 & 100\% & 100\% & 100\% & 100\% & 100\% & 100\% \\
\hline
\end{tabular}
\end{table}

\clearpage

\begin{figure}[ht]
    \begin{center}
    \includegraphics[width=1\columnwidth]{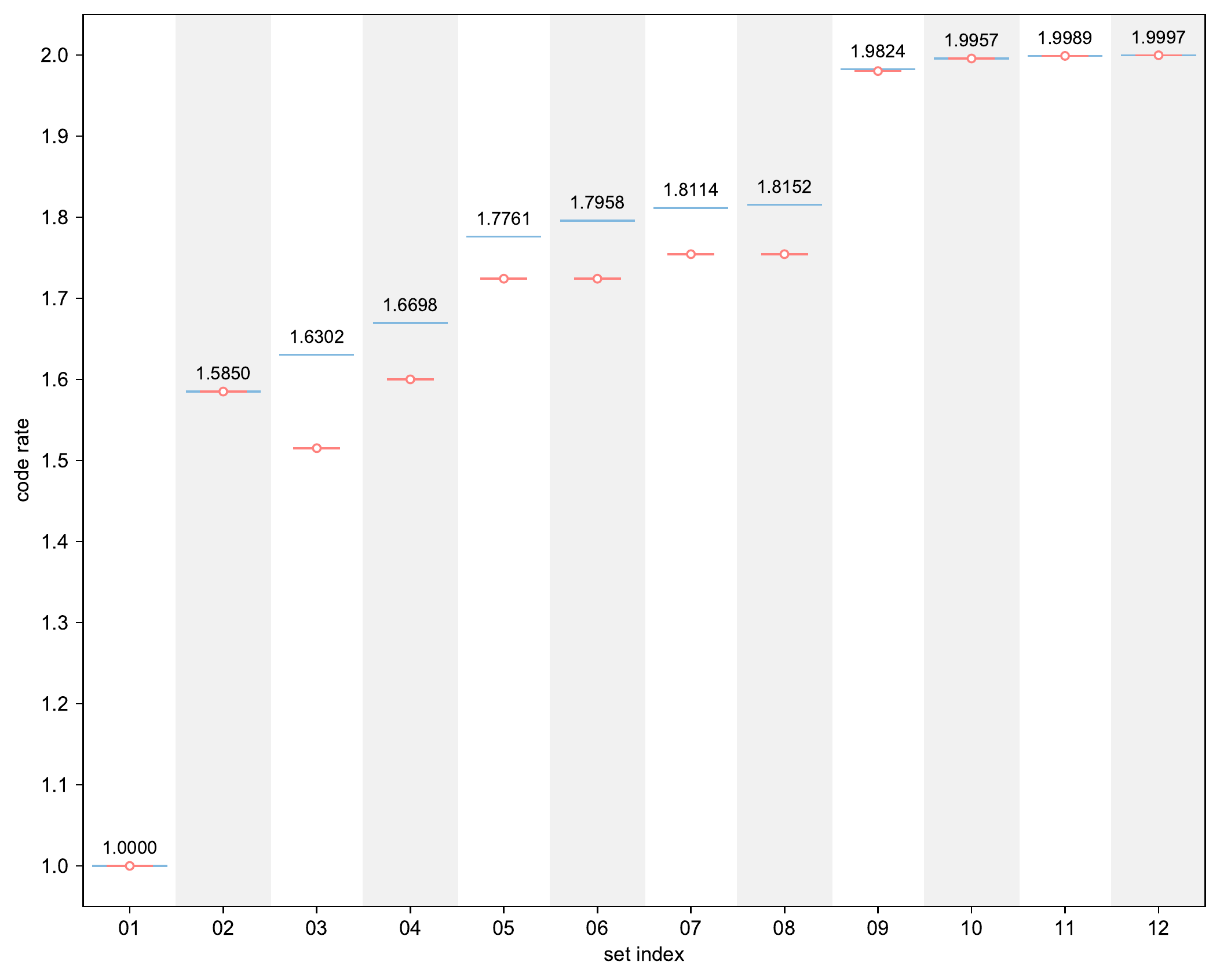}
    \caption{
    \textbf{Information density of graph-based coding algorithms versus their corresponding approximated capacity.}
    As the experimental setup, the length of binary message is 200 and the task repeat time is 100.
    In this experiment, the influence of payload length (or bit length) on information density is only considered, which does not include index range, error correcting range, and primer range.}
    \label{fig:code_rate}
    \end{center}
\end{figure}

\clearpage

\begin{figure}[ht]
    \begin{center}
    \includegraphics[width=1\columnwidth]{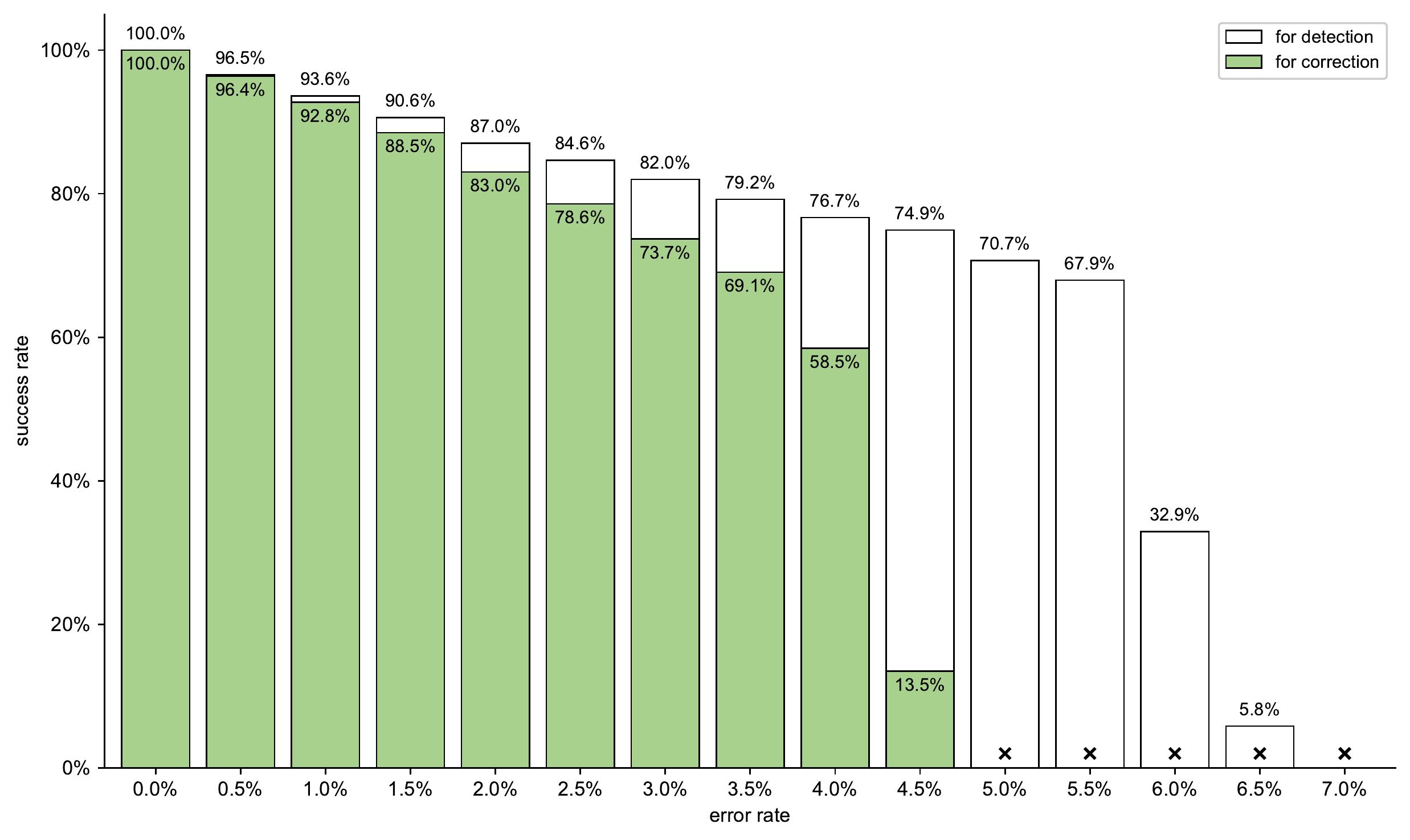}
    \caption{
    \textbf{Detection rate and correction rate under different error rates.}
    This is consistent with previous experiments (Figure~\ref{fig:repair}).
    The detection rate is usually much higher than the error correction rate.
    When the error rate of DNA sequences reaches $5.0\%$, SPIDER-WEB cannot correct all the errors or the number of solutions candidates exceeds $1,000$. 
    Meanwhile, SPIDER-WEB cannot detect all the errors when the error rate of DNA sequences reaches $7.5\%$.
    }
    \label{fig:detection}
    \end{center}
\end{figure}

\clearpage

\begin{figure}[ht]
    \begin{center}
    \includegraphics[width=1\columnwidth]{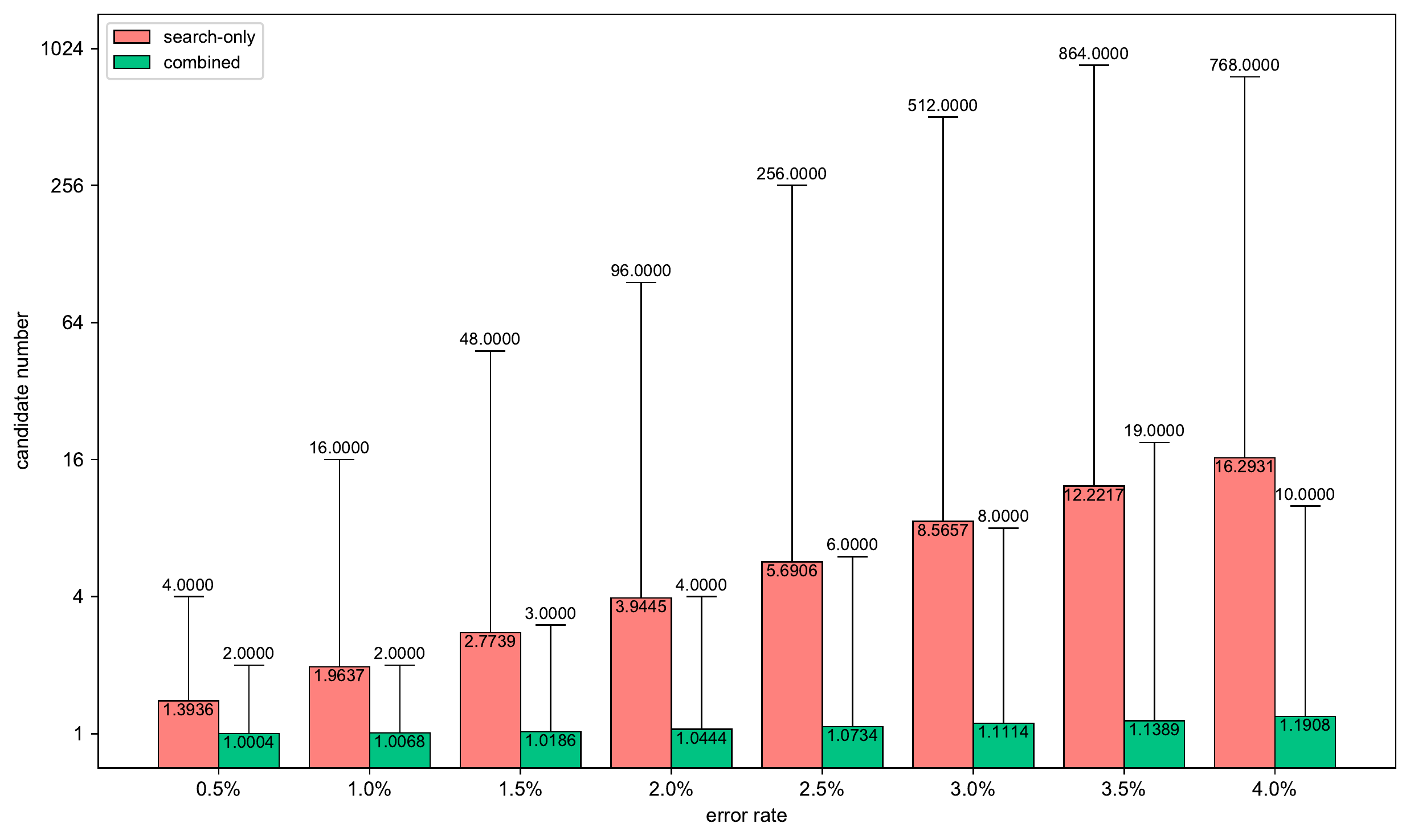}
    \caption{
    \textbf{Effect of Varshamov-Tenengolts path check.}
    Here, the word ``search-only'' refers to using local exhaustive reverse search alone. and the word ``combined'' represents that the candidates are the solution intersection of the local exhaustive reverse search and the path check. 
    The column chart describes the average value, while the top horizontal line describes the maximum value.}
    \label{fig:repair_reduce}
    \end{center}
\end{figure}

\clearpage

\begin{figure}[ht]
    \begin{center}
    \includegraphics[width=1\columnwidth]{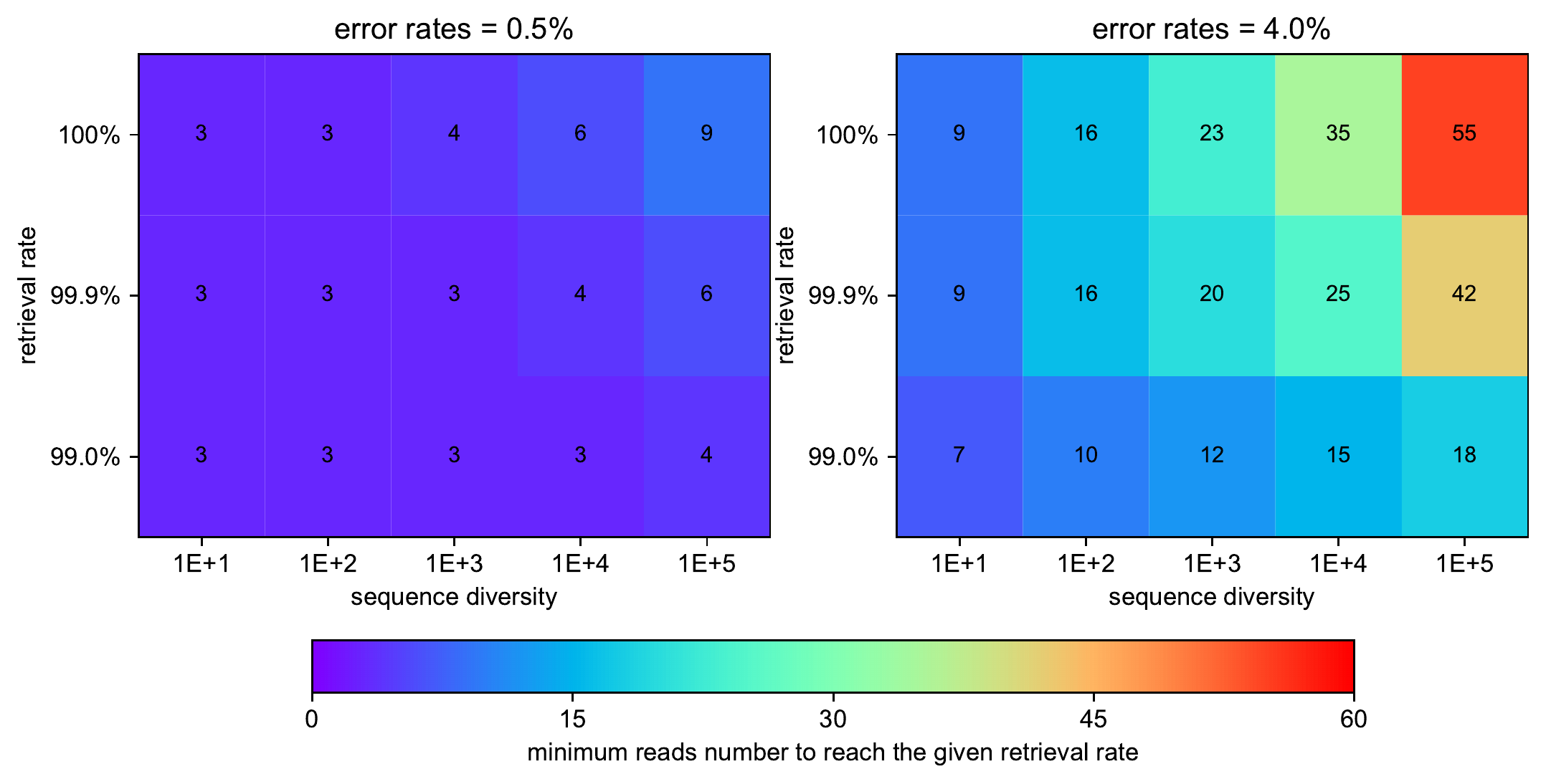}
    \caption{
    \textbf{Raw experiment data of minimum reads number under different error rates and retrieval rates.} 
    The implementation of this experiment is tentative.
    When $100$ experiments have reached or exceeded the given retrieval rate, the experiment can be allowed to end; otherwise, the reads number needs to be increased by one, and the experiment continue for $100$ times on this basis.}
    \label{fig:min_reads}
    \end{center}
\end{figure}

\clearpage

\begin{figure}[ht]
    \begin{center}
    \includegraphics[width=1\columnwidth]{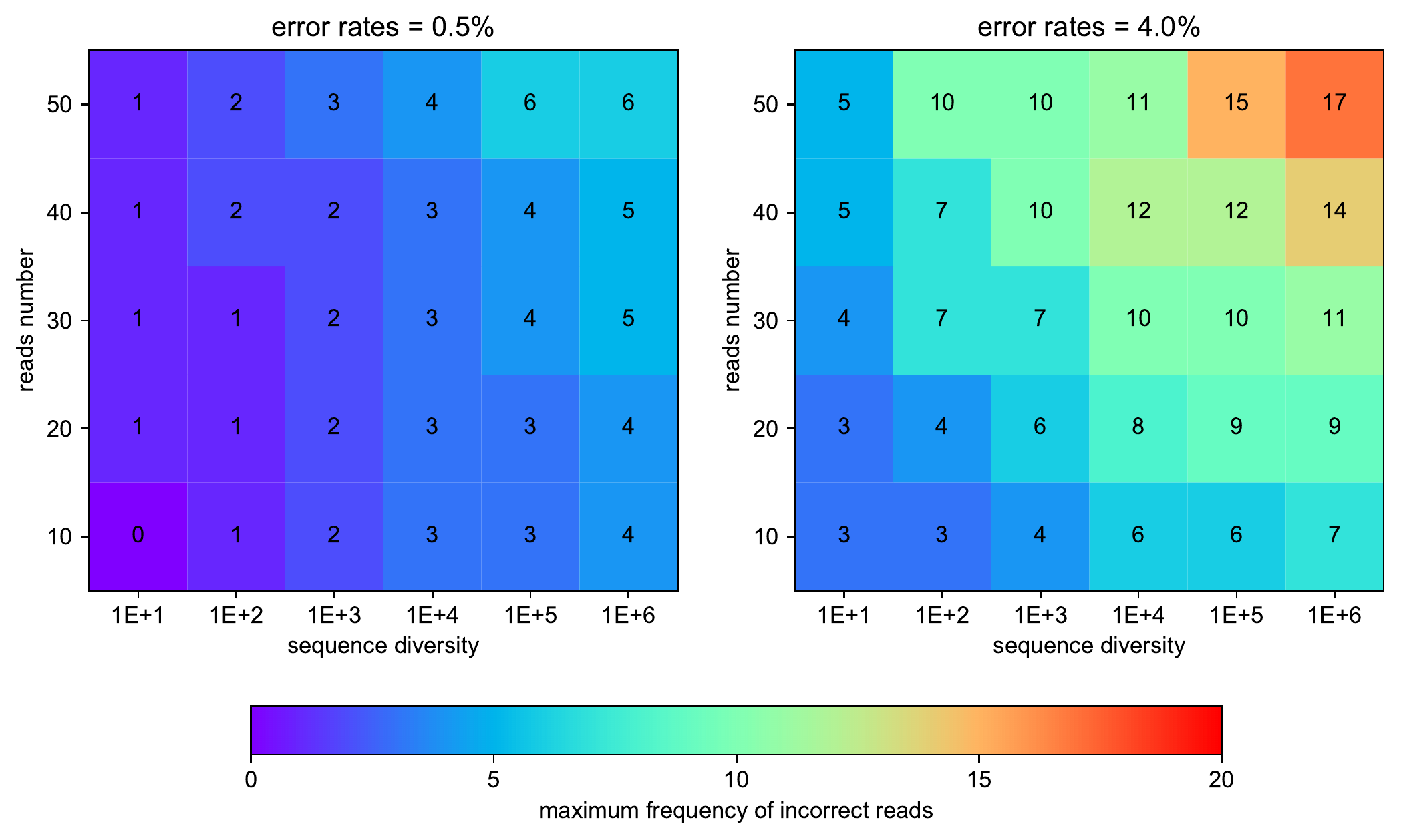}
    \caption{
    \textbf{Raw experiment data of maximum frequency of incorrect reads under different error rates and sequence diversities.} 
    $100$ random experiments are conducted under each setting.}
    \label{fig:min_threshold}
    \end{center}
\end{figure}

\clearpage

\begin{table}[htbp]
\centering
\caption{\textbf{Average vertex access frequency of local and global search under different error rates.}
The DNA sequence length is 200 nt and there are $1,000$ samples in each task.
The purpose of collecting the searching number of vertices rather than the actual time cost is to eliminate the uncertain impact of the implementation differences and programming languages.
The results are organized in ``average value (median value)''.
For global search~\cite{hart1968formal}, some samples have a large deviation at the initial stage, which makes their search steps abnormal. 
Therefore, the average value will be much greater than the median value.
}  
\label{tab:runtime}
\begin{tabular}{lll}
\hline
\textbf{error rate} & \textbf{local search} & \textbf{global search} \\
\hline
$0.0\%$ & $200.000$ ($200$) & $360637.693$ ($3338$) \\
$0.5\%$ & $227.449$ ($225$) & $328320.850$ ($3374$) \\
$1.0\%$ & $254.161$ ($251$) & $431748.949$ ($3500$) \\
$1.5\%$ & $277.872$ ($275$) & $329569.260$ ($3766$) \\
$2.0\%$ & $308.607$ ($307$) & $368018.044$ ($4576$) \\
$2.5\%$ & $334.428$ ($332$) & $295815.554$ ($4000$) \\
$3.0\%$ & $361.893$ ($358$) & $353709.747$ ($4566$) \\
$3.5\%$ & $388.530$ ($384$) & $334685.628$ ($4720$) \\
$4.0\%$ & $405.282$ ($402$) & $313815.952$ ($4986$) \\
\hline
\end{tabular}
\end{table}

\clearpage

\begin{figure}[ht]
    \begin{center}
    \includegraphics[width=1\columnwidth]{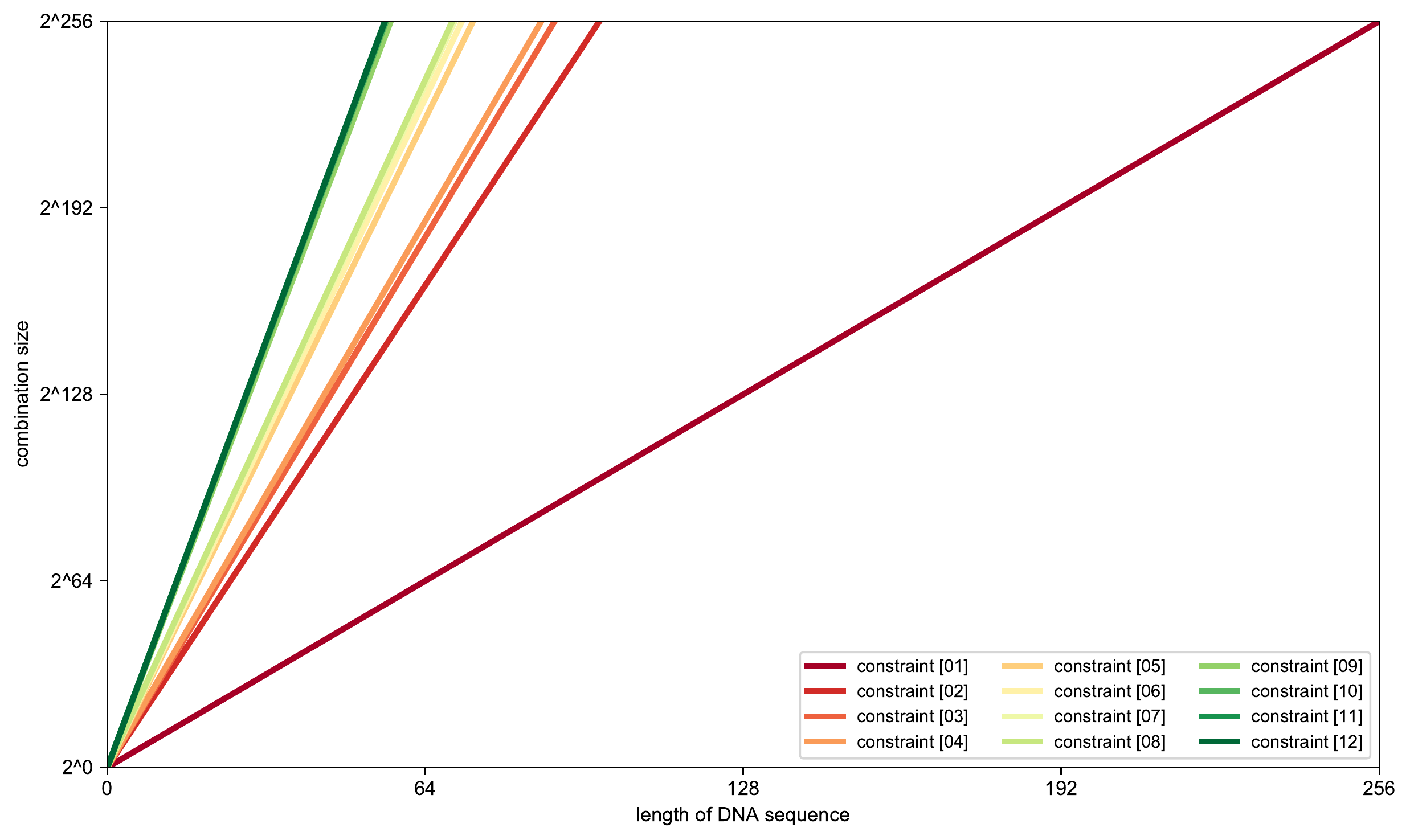}
    \caption{\textbf{Combination size under different constraints based on DNA sequence lengths.}
    Constraints are from Table~\ref{tab:constraint_sets}.}
    \label{fig:protection}
    \end{center}
\end{figure}

\clearpage

\begin{figure}[ht]
    \begin{center}
    \includegraphics[width=1\columnwidth]{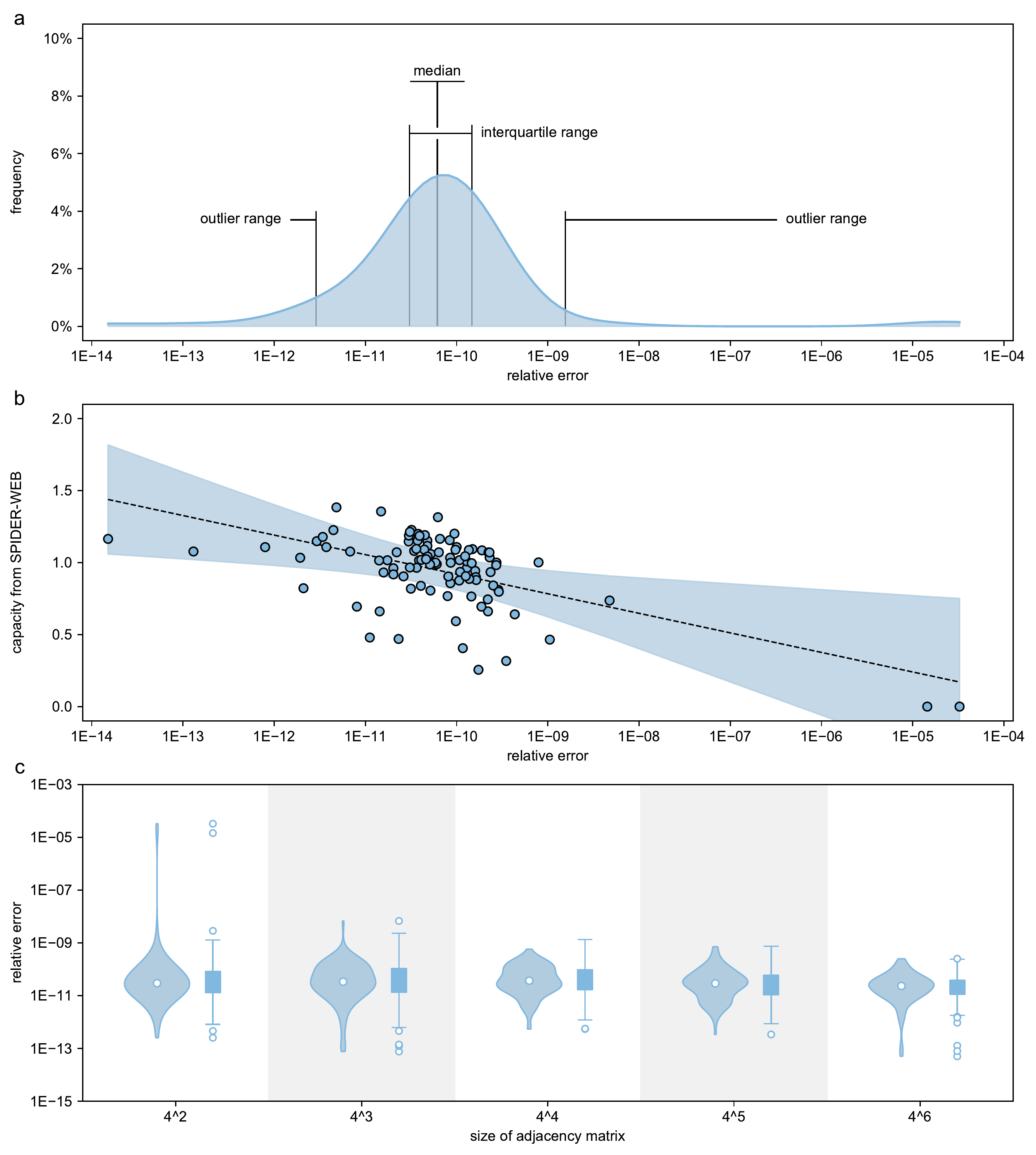}
    \caption{\textbf{Relative error statistics of the capacity approximation using random digraphs.}
    In (a) and (b), 100 random directed graphs are pruned from the graph without constraints.
    Among them, the median value is $2.97 \times 10^{-11}$, the lower and upper bound interquartile range are $1.30 \times 10^{-11}$ and $8.16 \times 10^{-11}$ respectively, the value of outliers is less than $8.26 \times 10^{-13}$ or more than $1.28 \times 10^{-9}$.
    For (c), 100 random directed graphs of different investigated observed lengths are pruned from the graph without constraints.
    The median values of each statistics are $2.97 \times 10^{-11}$, $3.37\times 10^{-11}$, $3.70 \times 10^{-11}$, $2.92 \times 10^{-11}$, and $2.33 \times 10^{-11}$, respectively.}
    \label{fig:capacity_evaluation}
    \end{center}
\end{figure}

\clearpage

\putbib[reference]
\label{sec:addition_reference}
\clearpage
\end{bibunit}
\end{document}